\documentclass[twocolumn,preprintnumbers,superscriptaddress,nofootinbib,aps,prd,floatfix]{revtex4}

\usepackage[utf8]{inputenc}

\usepackage{footmisc,enumerate}
\usepackage{multirow}
\usepackage{subfigure}
\usepackage{amsmath}
\usepackage{graphicx}
\usepackage{placeins}
\usepackage{xspace,slashed}
\usepackage{hyperref}
\hypersetup{colorlinks=true, citecolor=blue, urlcolor=blue, linkcolor=blue}
\usepackage[normalem]{ulem}
\usepackage{booktabs}

\renewcommand{\Ref}[1]{Ref.~\cite{#1}}
\newcommand{\Eq}[1]{Eq.~\eqref{#1}}
\newcommand{\Fig}[1]{Fig.~\ref{#1}}
\newcommand{\Tab}[1]{Tab.~\ref{#1}}
\newcommand{\Sec}[1]{Sec.~\ref{#1}}
\newcommand{\sw}{s_w}
\newcommand{\cw}{c_w}
\newcommand{\tw}{t_w}
\newcommand{\nn}{\nonumber}
\newcommand{\vev}{v}

\newcommand{\ndxs}[1]{\frac{1}{\sigma}\frac{{\rm d}\sigma}{{\rm d}#1}}
\renewcommand{\H}{{\mathcal{H}}}
\newcommand{\G}{{\mathcal{G}}}
\newcommand{\curr}[2]{c_{{#1}{#2}}\bar{#1}\gamma^\mu {#2}}
\def\OHq#1{\ensuremath{\mathcal{O}_{\varphi q}^{#1}}}
\def\OHu{\ensuremath{\mathcal{O}_{\varphi u}}}
\def\OHud{\ensuremath{\mathcal{O}_{\varphi ud}}}
\def\OHqW{\ensuremath{\mathcal{O}_{Hq}^W}}
\def\OHqZ{\ensuremath{\mathcal{O}_{Hq}^Z}}
\def\CHq#1{\ensuremath{C_{\varphi q}^{#1}}}
\def\CHu{\ensuremath{C_{\varphi u}}}
\def\CHud{\ensuremath{C_{\varphi ud}}}
\def\CHqW{\ensuremath{C_{\varphi q}^W}}
\def\CHqZ{\ensuremath{C_{\varphi q}^Z}}

\def\dzl{\delta_{Z,L}^t}
\def\dzr{\delta_{Z,R}^t}
\def\dwl{\delta_{W,L}}
\def\dwr{\delta_{W,R}}

\newcommand{\vp}{\varphi}
\newcommand{\tvp}{\widetilde{\varphi}}
\newcommand{\vpj }{\mbox{${\vp^\dag i\,\overleftrightarrow{D}_\mu\,\vp}$}}
\newcommand{\vpjt}{\mbox{${\vp^\dag i\,\overleftrightarrow{D}_\mu^{\,I}\,\vp}$}}
\begin{document}

\providecommand{\abs}[1]{\lvert#1\rvert}

\preprint{}

%%%%%%%%%%%%%%%%%%%%
\title{Electroweak Top Couplings, Partial Compositeness and Top Partner Searches}
%%%%%%%%%%%%%%%%%%%%

\begin{abstract}
Partial top quark compositeness is a crucial aspect of theories with strong electroweak symmetry breaking. Together with the heavy top partners that lift the top quark mass to its observed value, these theories predict correlated modifications of the top quark's electroweak couplings. Associated measurements therefore provide direct constraints on the ultraviolet structure of the underlying hypercolour dynamics. In this paper we employ a minimal version of top compositeness to discuss how measurements related to the top's electroweak gauge interactions can inform the potential composite nature of the TeV scale. In doing so, we identify the dominant factors that limit the BSM sensitivity. Extrapolating to a future 100 TeV hadron collider, we demonstrate that top quark measurements performed at highest precision can provide additional information to resonance search by performing a representative resonant top partner search that specifically targets the correlated resonant electroweak top partner signatures.
\end{abstract}

%%%%%%%%%%%%%%%%%%%%%%%%%%%%%%%%%%%%%%%%%%%%%%%%%%%%%%%%%%%%%%%%%
\author{Stephen Brown} \email{s.brown.7@research.gla.ac.uk}
\affiliation{School of Physics and Astronomy, University of Glasgow, Glasgow G12 8QQ, UK\\[0.1cm]}
%%%%
\author{Christoph Englert} \email{christoph.englert@glasgow.ac.uk}
\affiliation{School of Physics and Astronomy, University of Glasgow, Glasgow G12 8QQ, UK\\[0.1cm]}
%%%%
\author{Peter Galler} \email{peter.galler@glasgow.ac.uk}
\affiliation{School of Physics and Astronomy, University of Glasgow, Glasgow G12 8QQ, UK\\[0.1cm]}
%%%%
\author{Panagiotis Stylianou}\email{p.stylianou.1@research.gla.ac.uk} 
\affiliation{School of Physics and Astronomy, University of Glasgow, Glasgow G12 8QQ, UK\\[0.1cm]}
%%%%%%%%%%%%%%%%%%%%%%%%%%%%%%%%%%%%%%%%%%%%%%%%%%%%%%%%%%%%%%%%%

%%%%%%%%%%%%%%%%%%%%%%%
%%%%%%%%%%%%%%%%%%%%%%%

\maketitle

%%%%%%%%%%%%%%%%%%%%%%%%%%%%%%%%%%
\section{Introduction}
\label{sec:intro}
%%%%%%%%%%%%%%%%%%%%%%%%%%%%%%%%%%
Measurements at the Large Hadron Collider (LHC) have explored and constrained a range of 
ultraviolet (UV) completions of the Standard Model (SM) of Particle Physics. 
At the present stage of the LHC programme it is fair to say that unless new light, beyond the Standard Model (BSM) 
physics is hiding in experimentally challenging signatures, 
it is either weakly coupled to the SM or there is a considerable mass gap between the SM and the BSM spectrum. 
The latter avenue has motivated largely model-independent approaches based on effective field theory (EFT)
techniques recently. In case the SM's UV completion is both weakly coupled and scale separated to the extent 
that modifications of the low-energy SM Lagrangian become non-resolvable in the light of expected 
theoretical and experimental limitations, the EFT approach will become as challenged as measurements in the full model-context 
that the EFT can approximate. 
If, on the other hand, new physics is actually strongly coupled at larger energy scales, EFT-based methods
are suitable tools to capture the UV completions' dynamics and symmetry. Prime examples of such
theories are models with strong electroweak symmetry breaking~(EWSB, see~\cite{Contino:2010rs,Panico:2015jxa,Dawson:2018dcd,Cacciapaglia:2020kgq} for recent reviews). 

While the details of realistic UV models of compositeness vary in their microscopic structure,~e.g.~\cite{Ferretti:2013kya,Ferretti:2016upr}, they share common phenomenological aspects that are summarised in the so-called Minimal Composite Higgs Models (MCHMs)~\cite{Contino:2003ve,Agashe:2004rs,Contino:2006qr} (see also~\cite{Bellazzini:2014yua,Marzocca:2012zn,Pomarol:2012qf,Redi:2012ha}). This is possible as there are two necessary ingredients of pseudo-Nambu Goldstone Higgs theories: firstly, the explicit breaking of a global symmetry by weakly gauging a global (flavour) subgroup in the   
confining phase of a ``hypercolour'' interaction. Secondly, partial fermion compositeness~\cite{Terazawa:1976xx,Terazawa:1990mz,Kaplan:1991dc,Contino:2006nn} supplies an additional source of global symmetry breaking through (extended) hypercolour interactions. Both effects conspire to an effective low energy Higgs potential~\cite{Contino:2010rs,Contino:2003ve,Agashe:2004rs,Contino:2006qr,Ferretti:2016upr} 
of the form
\begin{equation}
V(h) = f^4 \left(  \beta \sin^2 {h\over f} - {\alpha+ 2\beta \over 4} \right)^2
\end{equation}
where $f$ is Goldstone boson decay constant, $h$ is a custodial isospin singlet for a given embedding of $SU(2)_L\times SU(2)_R$, and $\alpha,\beta$ are low energy constants (LECs) related to two- and four-point correlation functions of the (extended) hypercolour theory~\cite{Golterman:2015zwa,DelDebbio:2017ini}. The vacuum expectation value is determined as
\begin{equation}
\label{eq:ewsb}
\sin^2 {\langle h \rangle \over f} = {\alpha+ 2\beta \over 4\beta} = {v^2\over f^2}=  \xi
\end{equation}
where $\xi$ parametrises the model-dependent modifications of the physical Higgs boson to SM matter, see~e.g.~Ref.~\cite{Gillioz:2012se} for an overview. The physical Higgs mass is related to the LECs via
\begin{equation}
m_h^2 = f^2\left(8 \beta -2{\alpha^2\over \beta}\right).
\end{equation}

Symmetry breaking $\xi>0$ in Eq.~\eqref{eq:ewsb} constrains the LECs $\alpha,\beta \neq 0$, and experimental observations of the Higgs and electroweak bosons imply
\begin{equation}
\label{eq:cancel}
0.258 \simeq {m_h^2 \over v^2}  = 8(2\beta-\alpha)\,.
\end{equation}
This limits the parameter range that a realistic theory needs to reproduce. Furthermore, the region $\xi \ll 1$ which is required to have SM-like Higgs interactions as indicated by LHC measurements is accessed by $\alpha\simeq -2\beta$ which selects an isolated region in LEC parameter space. 

This is often understood as some indication of fine tuning, however, it can be shown that no linear combination of $\alpha,\beta$ is insensitive to four-point correlation functions~\cite{DelDebbio:2017ini}.
The computation of baryon four-point functions on the lattice is highly involved.\footnote{Progress has been made towards a better understanding of realistic composite Higgs theories using lattice simulations in Refs.~\cite{Hietanen:2014xca,DeGrand:2015lna,Ayyar:2017qdf,Ayyar:2018zuk,Ayyar:2018glg,DeGrand:2019vbx}.} Additional phenomenological input is needed to constrain concrete scenarios~\cite{DelDebbio:2017ini}, at least given the current status of lattice calculations. This also shows that there is technically no fine-tuning of the electroweak scale in these scenarios (yet), but an insufficient knowledge of the precise form of UV dynamics as can be expected from performing calculations in the interpolating hyperbaryon and meson picture.

Constraints or even the observation of partial compositeness in the top sector provide complementary phenomenological input and it is the purpose of this work to re-interpret existing LHC searches coherently along these lines. Extrapolating the current searches, we will also discuss the potential of the high-luminosity (HL-)LHC (13 TeV) and a future circular hadron-hadron collider (FCC-hh) to further narrow down the parameter space of strong interactions.

This paper is organised as follows:
In \Sec{sec:strongtop}, we review the basics of the composite top
scenario.
Our approach to constraining anomalous top couplings to $W$ and $Z$ bosons in
this model is outlined in \Sec{sec:topew}.
Following this strategy we discuss in \Sec{sec:results} the indirect
sensitivity reach of top measurements to coupling deformations as expected in
top compositeness theories at the LHC and also provide projections for a 100 TeV FCC-hh~\cite{Benedikt:2018csr} (see also~\cite{Agashe:2005vg,Kumar:2009vs}).
In \Sec{sec:reson}, we focus on a resonance search in a  representative $pp
\to T X, T \to t (Z \to \ell^+ \ell^-)$ final state, where $T$ is the top
partner and $X$ is either an additional $T$ or a third generation quark. This analysis
directly reflects the region where top-partial compositeness leads to new resonant structures as 
a consequence of modified weak top interactions. The
sensitivity of this direct search is compared with the indirect sensitivity
reach to demonstrate how top fits and concrete
resonance searches both contribute to a more detailed picture of top-partial compositeness at hadron colliders. Conclusions are given in \Sec{sec:conc}.

%%%%%%%%%%%%%%%%%%%%%%%%%%%%%%%%%%
\section{Strong coupling imprints in top quark interactions}
\label{sec:strongtop}
%%%%%%%%%%%%%%%%%%%%%%%%%%%%%%%%%%
Composite Higgs theories are conveniently expressed in terms of a Callen, Coleman, Wess, Zumino (CCWZ)
construction of Refs.~\cite{Coleman:1969sm,Callan:1969sn} (see also~\cite{Panico:2015jxa}) for a
given global symmetry breaking pattern $\G \to \H $. Denoting the $\G/ \H$ generators with $\hat T^A$ and those of $\H$ with $T^a$, the associated non-linear
sigma model field
\begin{equation}
\Sigma= \exp\{i 	\hat \phi^A \hat T^A/f\} \in \G\,
\end{equation}
captures the transformation properties of the (would-be) Goldstone bosons $\hat \phi^A$ under $g\in \G$ as
\begin{equation}
\label{eq:trafo}
\Sigma \to g \Sigma h^\dagger(g,\hat \phi)\,.
\end{equation}
From this, one can define kinetic terms by considering the $\G/\H$ part of 
\begin{equation}
\label{eq:oneforms}
\Sigma^\dagger 
\partial_\mu \Sigma = v^a_\mu T^a + p_\mu^{ A} \hat{T}^{ A} = v_\mu+ p_\mu\,,
\end{equation}
which transforms as $p_\mu \to h\hspace{0.04cm}p_\mu h^\dagger$~\cite{Callan:1969sn}. 
As indicated in Eq.~\eqref{eq:trafo}, this transformation will in general be non-linear as $h$ is $\hat \phi$ and $\G$-dependent, but
will reduce to linear transformations for $g \in \H$. If $\G/\H$ is a symmetric space, i.e. there
is an automorphism $A$: $A(T^a)=T^a$, $A(\hat T^A)= -\hat T^A$, we can consider a simplified object~\cite{Coleman:1969sm} 
\begin{equation}
U = \Sigma A(\Sigma)^\dagger 
\end{equation}
which lies in $\G/\H$ but transforms linearly under $\G$. 

For the purpose of this work we will consider a particular UV completion of MCHM5~\cite{Contino:2006qr}, which is based on
$SO(5)\to SO(4)$. Concrete ultraviolet completions of $\G=SO(5) \times U(1) \to SO(4) \times U(1)=\H$ require a larger symmetry, e.g. $SU(5)\to SO(5)\supset SO(4)$~\cite{Ferretti:2014qta,Golterman:2015zwa,Cacciapaglia:2015yra,Ferretti:2016upr,Golterman:2017vdj} and therefore typically lead to a richer pseudo-Nambu Goldstone boson and hyperbaryon phenomenology~\cite{DeGrand:2016htl,Ayyar:2017qdf,Belyaev:2016ftv,Englert:2016ktc,Ko:2016sht,DelDebbio:2017ini}. 
In this case the automorphism is related to complex conjugation and~\cite{Ferretti:2014qta} 
\begin{equation}
U=\Sigma \Sigma^T = \exp\{2 i \hat \phi^A \hat T^A/f\}
\end{equation}
with kinetic term
\begin{equation}
{\cal{L}}\supset {f^2\over 16} \text{Tr} ( \partial_\mu U \partial^\mu U^\dagger )\,.
\end{equation}
Weak gauging of a (sub)group of $H$ can be achieved in the lowest order in the Goldstone boson expansion by replacing the partial derivatives with covariant ones~\cite{Callan:1969sn,Giudice:2007fh}, from which we can derive Higgs interactions with weak gauge bosons. We will not explore this further in the following, but will assume this extension of MCHM5 to make contact with concrete UV extensions. Technically, this amounts to the underlying assumption of top partners being the lightest states in the TeV regime in this work when we will correlate the top partner masses with the top-electroweak coupling modifications in \Sec{sec:results}.

EWSB  in strongly coupled composite Higgs theories
relies on the presence of additional sources of global symmetry breaking as
weak gauging of the $SU(2)_L\times U(1)_Y$ will dynamically
align the vacuum in the symmetry-preserving direction.\footnote{While gauging 
QED in the pion sector leads to an excellent description of the $\pi^+,\pi^0$ mass splitting
QED remains exact. See~\cite{Contino:2006nn} for a detailed discussion of this instructive example.}
An elegant solution to this is partial compositeness~\cite{Kaplan:1991dc,Contino:2006nn,Sannino:2016sfx,Cacciapaglia:2017cdi}. 
Partial compositeness traces the fermion mass hierarchy to mixing of massless elementary fermions with
composite hyperbaryons of the strong interactions. This not only serves the purpose of misaligning the vacuum from 
the $SU(2)_L\times U(1)_Y$ direction, rendering the Higgs a pseudo-Nambu Goldstone
boson, but in parallel lifts the top and bottom masses to their observed values. Phenomenologically,
this results in a tight correlation of top and Higgs interactions, which is a non-perturbative example of
the close relation of the Higgs and top-quark interactions in generic BSM theories.

A minimal effective Lagrangian of partial compositeness in the light of $Z\bar b_L b_L$ coupling constraints~\cite{Contino:2006qr} is given by a scenario based on $SU(5)/SO(5)$~\cite{Ferretti:2014qta} (which again resembles the SO(5)/SO(4) pattern with symmetric mass terms)
\begin{multline}
\label{eq:partcomp}
	-{\cal{L}} \supset {M} \bar\Psi \Psi + \lambda_q f \bar{\hat{Q}}_L\Sigma \Psi_R + \lambda_t f \bar{\hat{t}}_R \Sigma^\ast\Psi_L \\ +
	\sqrt{2} \mu_b \text{Tr}( \bar{\hat{Q}}_L U \hat{b}_R ) + \text{h.c.}\,.
\end{multline}
$\Psi$ represents the vector-like composite baryons in the low energy effective theory that form a $\bf{5}$ of
$SO(5)$ and transform in the fundamental representation of $SU(3)_C$
\begin{alignat}{5}
  \label{eq:tp}
    \Psi      &= \frac{1}{\sqrt{2}} \left( \begin{matrix} iB - iX \\ B+X \\ iT+iY \\ -T+Y \\\sqrt{2}iR \end{matrix} \right)\,.
\end{alignat}
$\Psi$ decomposes into a bi-doublet and a singlet under $SU(2)_L\times SU(2)_R$~\cite{DeSimone:2012fs} thus implementing the custodial $SU(2)$ mechanism of Ref.~\cite{Agashe:2006at}. Under the SM gauge interactions $SU(3)_C\times SU(2)_L\times U(1)_Y$, these fields transform as
$(T,B) \in (\mathbf{3},\mathbf{2})_{1/6}$,
$R \in (\mathbf{3},\mathbf{1})_{2/3}$, and
$(X,Y) \in (\mathbf{3},\mathbf{2})_{7/6}$.
$\hat{Q}_L\supset (t_L,b_L)$, $\hat{t}_R\supset t_R$, and $\hat{b}_R\supset b_R$ are $SO(5)$ spurions
\begin{equation}
\hat{Q}_L = 
\left(\begin{matrix} i b_L \\ b_L \\i t_L \\ -t_L \\ 0 \end{matrix} \right)\,,
\quad 
\hat{t}_R = \left(\begin{matrix} 0 \\ 0 \\ 0  \\ 0 \\ t_R \end{matrix} \right)\,,
\quad
\hat{b}_R =  \left(\begin{matrix} 0 \\ 0 \\ 0  \\ 0 \\ b_R \end{matrix} \right) \,.
\end{equation}
This additional source of $SO(5)$ breaking leads to EWSB as it implies a finite contribution to effective Higgs potential, and lifts the top mass $\sim f \lambda_q\lambda_t v /M$ in the large $M$ limit.
We can expand the Lagrangian of Eq.~\eqref{eq:partcomp} to obtain the top partner mass mixing
\begin{equation}
  \label{eq:topmass}
  {\cal{M}}_T=\left( \begin{matrix} 0  & {\lambda_q\over 2} f  (1+c_h)  & {\lambda_q\over 2} f (1-c_h) & {\lambda_q\over \sqrt{2} } f s_h\\
      {\lambda_t\over\sqrt{2}} f s_h & M & 0 & 0 \\
      -{\lambda_t\over\sqrt{2}} f s_h	& 0 & M & 0 \\
      \lambda_t f c_h & 0 & 0 & M
    \end{matrix}\right)\, ,
\end{equation}
where $c_h=\cos (h/f)$ and $s_h=\sin (h/f)$. Expanding $c_h,s_h$ around $\langle h \rangle$ gives rise to the Higgs-top (partner) interactions.
The mass mixing in the bottom sector reads
\begin{equation}
  \label{eq:bottmass}
  {\cal{M}}_B=\\\left( \begin{matrix} \mu_bs_hc_h  & \lambda_q f  \\ 0 & M
 \end{matrix}\right) \,.
\end{equation}

The mass eigenstates are obtained through bi-unitary transformations, which modify the weak and Higgs couplings of the physical top and bottom quarks compared to the SM by ``rotating in'' some of the top and bottom partner's weak interaction currents\footnote{Similar correlations are observed in models that target dark matter and B anomalies, see Ref.~\cite{Cline:2017aed}.} (following the notation of~\cite{Ferretti:2014qta})
\begin{equation}
{\cal{L}}\supset \bar \Psi \gamma^\mu \left( {2\over 3} e A_\mu  -{2\over 3}\tw e Z_\mu + v_\mu  + Kp_\mu\right)\Psi
\end{equation}
with $v_\mu,p_\mu$ arising from Eq.~\eqref{eq:oneforms} after gauging. $K$ is an additional undetermined LEC. This leads to currents
\begin{subequations}
\label{eq:jcurr}
\begin{multline}
\label{eq:zcurr}
J^\mu_Z/e= \curr{X}{X} + \curr{T}{T} + \curr{Y}{Y} \\ + \curr{R}{R}
+\curr{B}{B} + (\curr{R}{T} + \text{h.c.}) \\+ (\curr{R}{Y} + \text{h.c.}) 
+(\curr{T}{Y} + \text{h.c.})
\end{multline}
and 
\begin{multline}
\label{eq:wcurr}
J^\mu_{W^+}/e= \curr{X}{T} + \curr{X}{Y} + \curr{X}{R}\\ + \curr{T}{B} + \curr{Y}{B} + \curr{R}{B}\,,
\end{multline}
\end{subequations}
with coefficients $c_{i}$ 
\begin{equation}
\begin{split}
c_{XX} =& {1\over \sw\cw} \left({1\over 2} - {5\over 3} \sw^2 \right) \\
c_{TT} =&  -{2\over 3} t_w + {c_h\over 2 \sw\cw} \\
c_{YY} =&  -{2\over 3} t_w -{c_h\over 2\sw\cw}    \\
c_{RR} =& -{2\over 3} t_w  \\
c_{BB} =&  {1\over \sw\cw} \left(-{1\over 2} + {1\over 3} \sw^2 \right) \\
c_{TY} =&  0 \\
c_{RT} = c_{RY} =&  K {s_h\over 2\sqrt{2} \sw\cw}\,.
\end{split}
\end{equation}
Similarly, the $W$ couplings are
\begin{equation}
\begin{split}
c_{XT} = c_{YB} = &  {1-c_h \over 2\sqrt{2}\sw} \\
c_{XY} = c_{TB} = &  {1+c_h \over 2\sqrt{2}\sw} \\
c_{RB} = -c_{XR} = &  K {s_h\over 2 \sw}  \,.
\end{split}
\end{equation}
$\sw,\cw,\tw$ are the sine, cosine and tangent of the Weinberg angle, respectively. 

A non-vanishing $K$ significantly alters the tight correlation of the top
partner mass and coupling modifications of the top due to the mixing with heavy top partners.
In case $K=0$, a small top partner mass has to be compensated by a large mixing between top and top partners
in order to lift the mass of the elementary top to its physically observed value. The mixing angle
in turn determines the electroweak coupling deviations of the top quark in the mass eigenbasis. Hence,
for $K=0$ there exists a strong correlation between top partner mass and top coupling deviation.
However, if $K$ is allowed to take values $K\neq 0$ this correlation is loosened which in parallel
opens up momentum enhanced decays $T\to ht$~\cite{Ferretti:2014qta}.
In \Sec{sec:results} we study the dependence of the sensitivity on the parameter $K$ in indirect searches and
use this information to discuss the sensitivity gap with direct searches in \Sec{sec:reson}.

In addition to the coupling modifications of the top-associated currents, amplitudes receive corrections
from propagating top partners, for which we provide a short EFT analysis in appendix~\ref{app:eft} up to mass dimension eight. In the mass
basis these propagating degrees of freedom generate ``genuine'' higher dimensional effects and are therefore suppressed 
compared to the dimension four top-coupling modifications. Working with a concrete UV scenario, we have directly verified this suppression using a full simulation of propagating top partners in the limit where they are not resolved as resonances. We therefore neglect these contributions in our coupling analysis, but return to the relevance of resonance searches in \Sec{sec:reson}.

%%%%%%%%%%%%%%%%%%%%%%%%%%%%
\begin{table*}[t!]
\begin{center}
\caption{Experimental analyses used to determine constraints on anomalous top quark couplings. $tjZ$ denotes single-top t-channel production in association with a $Z$ boson.}
\label{tab:analyses}
{\begin{footnotesize}
{\renewcommand{\arraystretch}{1.6}
\renewcommand{\tabcolsep}{0.13cm}
\begin{tabular}[t]{lllll}
\hline\hline
Analysis & Collaboration & $\sqrt{s}$ [TeV] & Observables & dof\\ 
\hline
\multicolumn{5}{l}{single top $t$-channel}\\
\hline
1503.05027~\cite{Aaltonen:2015cra} & CDF, D0 & 1.96 & $\sigma_{\rm tot}$ & 1\\
1406.7844~\cite{Aad:2014fwa} & ATLAS & 7 &
    $\frac{\sigma_t}{\sigma_{\bar t}}$, & 1\\[-0.15cm]
&&& $\ndxs{p_\perp^t}$, $\ndxs{p_\perp^{\bar t}}$, & 8\\[-0.15cm]
&&& $\ndxs{|y_t|}$, $\ndxs{|y_{\bar t}|}$ & 6\\
1902.07158~\cite{Aaboud:2019pkc} & ATLAS,CMS & 7,8 & $\sigma_{\rm tot}$ & 2\\
1609.03920~\cite{Aaboud:2016ymp} & ATLAS & 13 & $\sigma_t$, $\frac{\sigma_t}{\sigma_{\bar t}}$ & 2\\
1812.10514~\cite{Sirunyan:2018rlu} & CMS & 13 & $\frac{\sigma_t}{\sigma_{\bar t}}$, $\sigma_t$ & 2\\
\hline
\multicolumn{5}{l}{single top $s$-channel}\\
\hline
1402.5126~\cite{CDF:2014uma} & CDF, D0 & 1.96 & $\sigma_{\rm tot}$ & 1\\
1902.07158~\cite{Aaboud:2019pkc} & ATLAS, CMS & 7, 8 & $\sigma_{\rm tot}$ & 2\\
\hline
\multicolumn{5}{l}{$tW$}\\
\hline
1902.07158~\cite{Aaboud:2019pkc} & ATLAS, CMS & 7, 8 & $\sigma_{\rm tot}$ & 2\\ 
1612.07231~\cite{Aaboud:2016lpj} & ATLAS & 13 & $\sigma_{\rm tot}$ & 1\\
1805.07399~\cite{Sirunyan:2018lcp} & CMS & 13 & $\sigma_{\rm tot}$ & 1\\
\hline
\multicolumn{5}{l}{$tjZ$}\\
\hline
1710.03659~\cite{Aaboud:2017ylb} & ATLAS & 13 & $\sigma_{\rm tot}$ & 1\\
1812.05900~\cite{Sirunyan:2018zgs} & CMS & 13 & $\sigma_{\rm tot}$ & 1\\
\hline
\hline
\end{tabular}}
\end{footnotesize}}
\hspace{-0.3cm}
{\footnotesize
{\renewcommand{\arraystretch}{1.6}
\renewcommand{\tabcolsep}{0.13cm}
\begin{tabular}[t]{lllll}
\hline\hline
Analysis & Collaboration & $\sqrt{s}$ [TeV] & Observables & dof\\ 
\hline
\multicolumn{5}{l}{$t\bar tZ$}\\
\hline
1509.05276~\cite{Aad:2015eua} & ATLAS & 8 & $\sigma_{\rm tot}$ & 1\\
1510.01131~\cite{Khachatryan:2015sha} & CMS & 8 & $\sigma_{\rm tot}$ & 1\\
1901.03584~\cite{Aaboud:2019njj} & ATLAS & 13 & $\sigma_{\rm tot}$ & 1\\
1907.11270~\cite{CMS:2019too} & CMS & 13 & $\sigma_{\rm tot}$, $\ndxs{p_\perp^Z}$, & 4\\[-0.15cm]
&&&$\ndxs{\cos\theta^*_Z}$ & 3\\
\hline
\multicolumn{5}{l}{$W$ boson helicity fractions}\\
\hline
1211.4523~\cite{Aaltonen:2012lua} & CDF & 1.96 & $F_0$, $F_R$ & 2\\
1205.2484~\cite{Aad:2012ky} & ATLAS & 7 & $F_0$, $F_L$, $F_R$ & 3\\
1308.3879~\cite{Chatrchyan:2013jna} & CMS & 7 & $F_0$, $F_L$, $F_R$ & 3\\
1612.02577~\cite{Aaboud:2016hsq} & ATLAS & 8 & $F_0$, $F_L$ & 2\\
\hline
\multicolumn{5}{l}{top quark decay width}\\
\hline
1201.4156~\cite{Abazov:2012vd} & D0 & 1.96 & $\Gamma_t$ & 1\\
1308.4050~\cite{Aaltonen:2013kna} & CDF & 1.96 & $\Gamma_t$ & 1\\
1709.04207~\cite{Aaboud:2017uqq} & ATLAS & 8 & $\Gamma_t$ & 1\\
\hline
\hline
\end{tabular}}}
\end{center}
\end{table*}
%%%%%%%%%%%%%%%%%%%%%%%%%%%%

%%%%%%%%%%%%%%%%%%%%%%%%%%%%%%%%%%
\section{Electroweak Top Property Constraints}
\label{sec:topew}
%%%%%%%%%%%%%%%%%%%%%%%%%%%%%%%%%
The weak couplings of the SM top and bottom quarks are modified due to the
mixing with the top and bottom partners in the mass eigenbasis. In particular,
these are modifications of the left and right-handed vectorial couplings to the $W$ and
$Z$ bosons which can be parametrised as follows
\begin{eqnarray}
\mathcal{L} & \supset & \bar{t}\gamma^{\mu}\left[g^t_LP_L+g^t_RP_R\right]tZ_{\mu}\nn\\
& + & \bar{b}\gamma^{\mu}\left[g^b_LP_L+g^b_RP_R\right]bZ_{\mu}\nn\\
& + & \left(\bar{b}\gamma^{\mu}\left[V_LP_L+V_RP_R\right]tW^+_{\mu} + {\rm h.c.}\right)\,.
\label{eq:WZcoupling}
\end{eqnarray}
The anomalous couplings of the top quark, i.e. the relative deviation with respect to the SM, are denoted by~$\delta$
\begin{eqnarray}
g^t_L
& = & -\frac{g}{2\cos\theta_W}\left(1-\frac{4}{3}\sin^2\theta_W\right)\Big[1+\dzl\Big]\,,
\label{eq:gtL}\\
g^t_R
& = & \frac{2g\sin^2\theta_W}{3\cos\theta_W}\Big[1+\dzr\Big]\,,
\label{eq:gtR}\\
V_L
& = & -\frac{g}{\sqrt{2}}\Big[1+\dwl\Big]\,,
\label{eq:VL}\\
V_R
& = & -\frac{g}{\sqrt{2}}\dwr\,,
\label{eq:VR}
\end{eqnarray}
where $g$ is the weak coupling constant associated with the $SU(2)_L$ gauge group
and $\theta_W$ is the Weinberg angle. Note that $\dwr$ is normalised
to the left-handed SM coupling of the top quark to the $W$ boson. Technically, we
implement the anomalous couplings in terms of Wilson coefficients in an effective
Lagrangian of dimension six operators. The relation between the $\delta$ parameters
and the Wilson coefficients in the Warsaw basis~\cite{Grzadkowski:2010es} is given
in appendix~\ref{app:anomalous} . The parametrisation in terms of Wilson coefficients
allows us to use an updated version of the \textsc{TopFitter} frame work (which will be described in detail
elsewhere~\cite{toapp}) to obtain constraints on the anomalous couplings of the top quark.
The anomalous couplings of bottom quarks to $Z$ bosons are phenomenologically less relevant by construction~\cite{Contino:2006qr}.

We obtain constraints on the anomalous couplings by comparing them to
experimental results for observables that are sensitive to the vectorial weak
couplings of the top quark. Specifically, we include in the fit 21 experimental
analyses~\cite{
Abazov:2012vd,		%1201.4156
Aad:2012ky,		%1205.2484
Aaltonen:2012lua,	%1211.4523
Chatrchyan:2013jna,	%1308.3879
Aaltonen:2013kna,	%1308.4050
CDF:2014uma,		%1402.5126
Aad:2014fwa,		%1406.7844
Aaltonen:2015cra,	%1503.05027
Aad:2015eua,		%1509.05276
Khachatryan:2015sha,	%1510.01131
Aaboud:2016ymp,		%1609.03920
Aaboud:2016hsq,		%1612.02577
Aaboud:2016lpj,		%1612.07231
Aaboud:2017uqq,		%1709.04207
Aaboud:2017ylb,		%1710.03659
Sirunyan:2018lcp,	%1805.07399
Sirunyan:2018zgs,	%1812.05900
Sirunyan:2018rlu,	%1812.10514
Aaboud:2019njj,		%1901.03584
Aaboud:2019pkc,		%1902.07158
CMS:2019too},		%1907.11270
which are presented in \Tab{tab:analyses} and amount to a total of $N=54$ degrees of freedom.

The likelihood
provided by \textsc{TopFitter} is defined as
\begin{multline}
-2\log L(\boldsymbol{\delta})\\
\qquad=\sum_{i,j=1}^{N}\left(X_i^{\rm exp}-X_i^{\rm th}(\boldsymbol{\delta})\right)(V^{-1})_{ij}\left(X_j^{\rm exp}-X_j^{\rm th}(\boldsymbol{\delta})\right)\,,
\label{eq:logL}
\end{multline}
where $X_i^{\rm exp}$ is the experimental result for the observable $X_i$ and
$X_i^{\rm th}(\boldsymbol{\delta})$ is the theoretical prediction which depends on the anomalous couplings $\dzl$,
$\dzr$, $\dwl$ and $\dwr$ collectively denoted by
$\boldsymbol{\delta}$. The inverse covariance matrix is denoted by $V^{-1}$ and
takes into account bin-to-bin correlations provided by the experimental
collaborations.
The theoretical uncertainties result from independently varying renormalisation
and factorization scale $\mu_R,\mu_F=\{m_t/2,m_t,2m_t\}$\footnote{$m_t$ denotes
the top quark mass and is set to $m_t=172.5$ GeV in alignment with the value used in
the experimental analyses in \Tab{tab:analyses}.}. Furthermore, we take
uncertainties on the parton distribution functions (PDF) and the strong coupling constant
$\alpha_s$ into account and evaluate
them according to the PDF4LHC recommendations~\cite{Butterworth:2015oua} using the
\texttt{PDF4LHC15\_nlo\_30\_pdfas} PDF set. Experimental, scale, PDF and $\alpha_s$ uncertainties
are added in quadrature. 

The SM contribution to the observable predictions $X_i^{\rm th}$
is computed at next-to-leading order QCD. The contributions from the anomalous couplings
are computed at leading order owing to the fact that we scan over small values for the
anomalous couplings and ignore additional contributions to the strong corrections. We take into account contributions that are quadratic and bilinear
in the anomalous couplings but have verified that they have only a small effect on the likelihood.

The theoretical predictions for both SM and anomalous couplings are obtained from \mbox{\textsc{MadGraph5\_aMC@NLO}} \cite{Alwall:2011uj,Alwall:2014hca} which
is the Monte Carlo generator used by \textsc{TopFitter}. The anomalous couplings are mapped to Wilson coefficients in the SM effective field theory (see appendix~\ref{app:anomalous}) and theoretical predictions are evaluated using the \mbox{\textsc{SmeftSim}}~\cite{Brivio:2017btx} UFO~\cite{Degrande:2011ua} model.
A parton shower and detector simulation is not necessary since the experimental results in \Tab{tab:analyses} are unfolded to parton level.

The likelihood in \Eq{eq:logL} is used to exclude anomalous couplings at a confidence level (CL) of 95\%.
A point $\boldsymbol{\delta}$ in the parameter space of the anomalous couplings is considered excluded if
\begin{equation}
1-{\rm CL}>\int_{-2\log L(\boldsymbol{\delta})}^{\infty}\hbox{d}x\,f_{\chi^2}(x,k)\,,
\label{eq:CL}
\end{equation}
where $f_{\chi^2}(x,k)$ is the $\chi^2$ probability distribution and \mbox{$k=N$} is the number of degrees of freedom.

Partial compositeness imposes strong correlations between the different anomalous couplings. Hence,
individual or marginalised bounds are not applicable since they would neglect these
correlations and lead to incorrect exclusions.
Instead, we scan over the model's parameter space and calculate the anomalous top couplings
that correspond to each sample point. We determine whether the
parameter points are excluded at 95\% confidence based on \Eq{eq:CL} using the likelihood
in \Eq{eq:logL} which includes the experimental input in
\Tab{tab:analyses} and is implemented by \textsc{TopFitter}. This
procedure takes the correlations between the anomalous couplings into account because the scan is performed in the
parameter space of the underlying model and then mapped to the weak vectorial top couplings.

In the next section we give details about the parameter scan and present the results
contrasting the current experimental situation with projections to larger integrated
luminosities and future colliders.

%%%%%%%%%%%%%%%%%%%%%%%%%%%%%%%%%%
\begin{figure*}[!t]
\begin{center}
\includegraphics[width=0.45\textwidth]{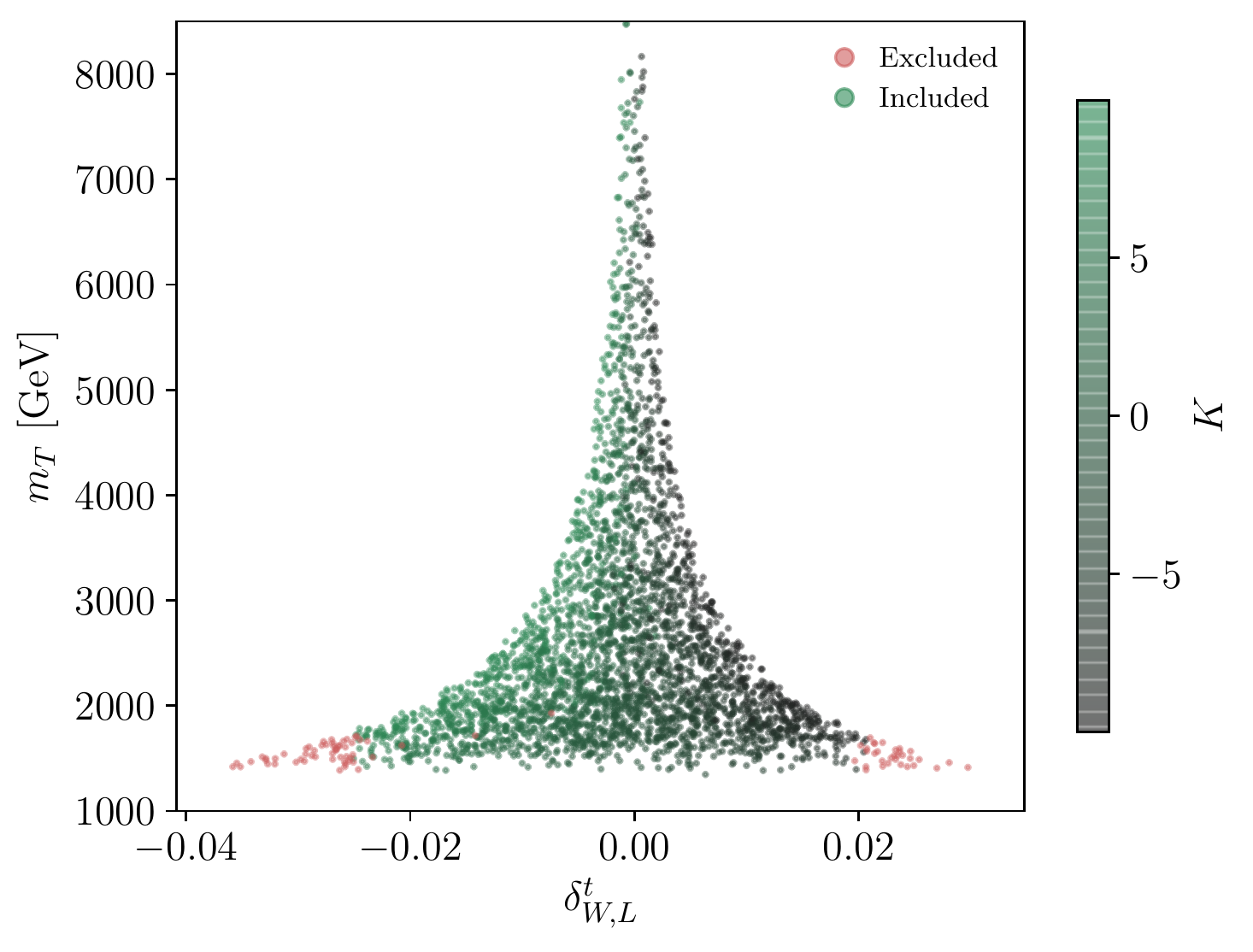}
\hskip 0.5cm
\includegraphics[width=0.45\textwidth]{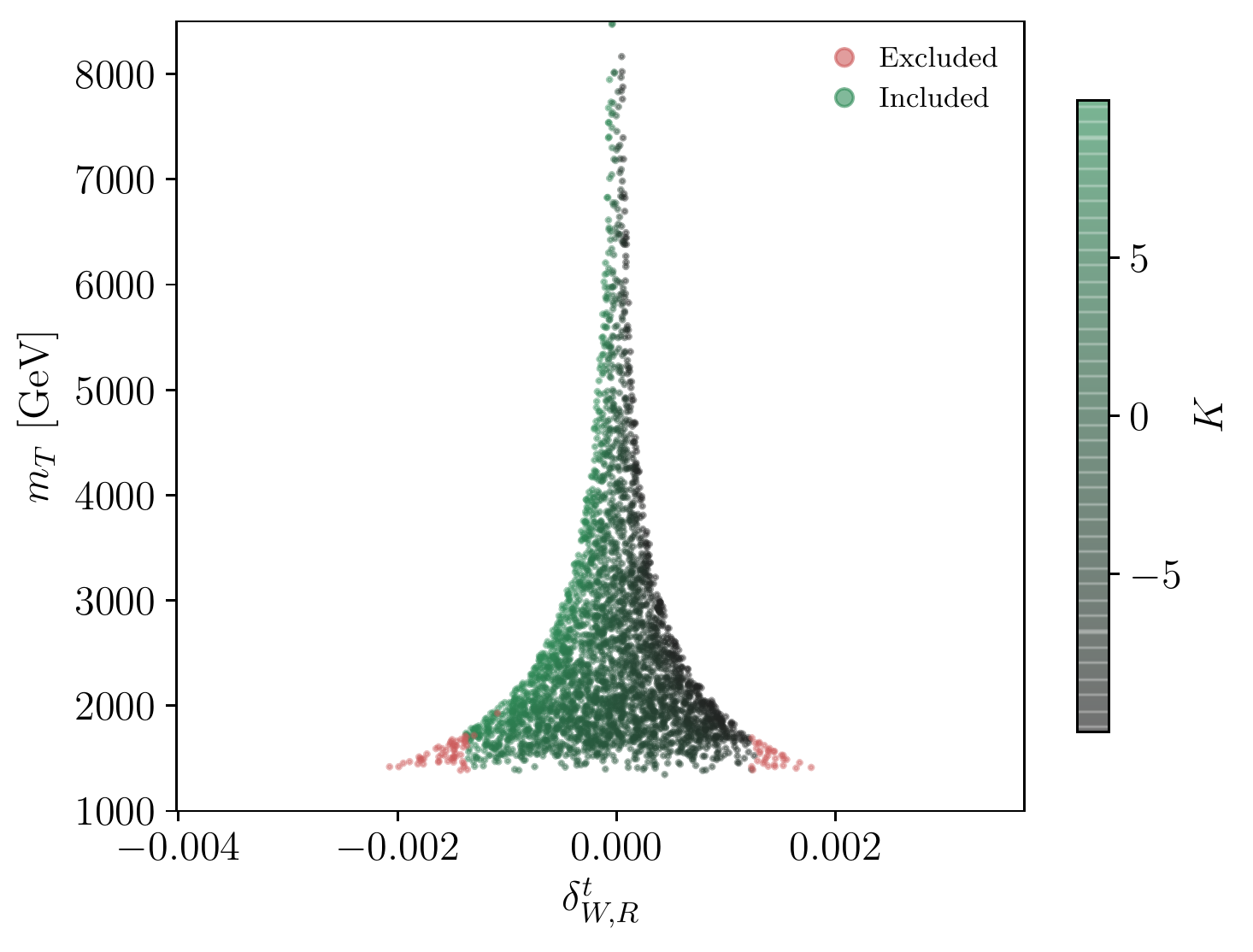}\\[0.3cm]
\includegraphics[width=0.45\textwidth]{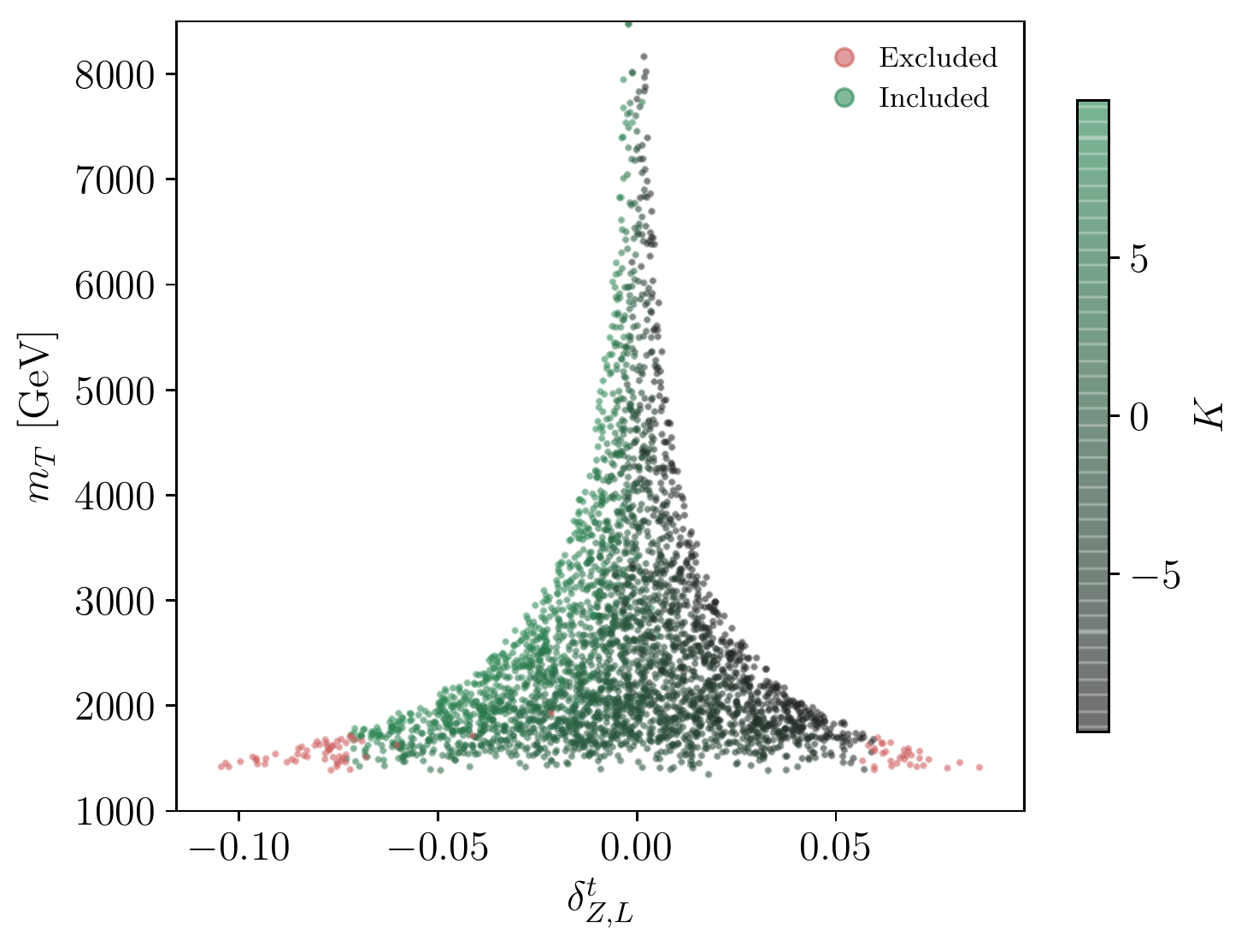}
\hskip 0.5cm
\includegraphics[width=0.45\textwidth]{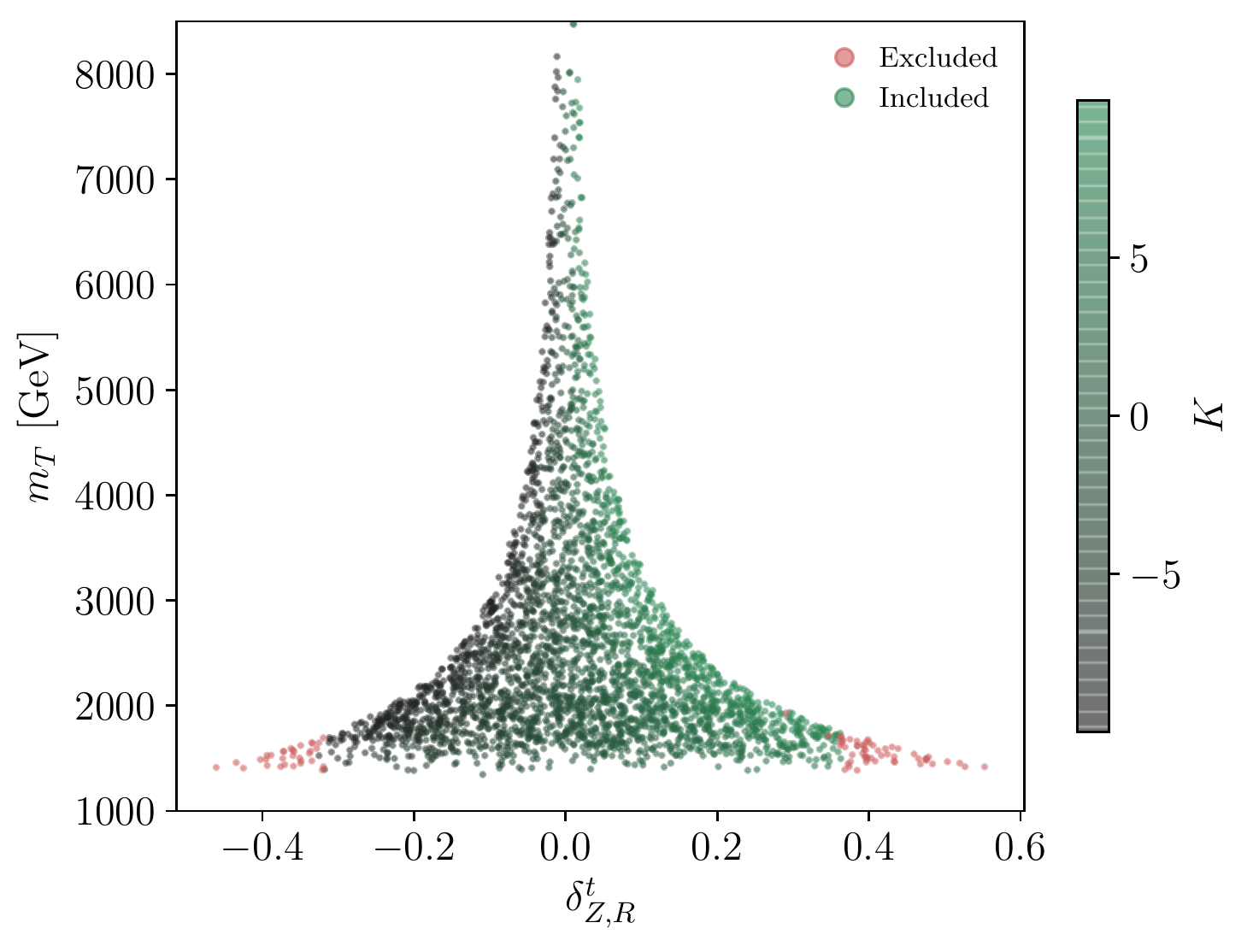}
\caption{Correlation between top partner mass $m_T$ and anomalous top quark couplings in the light of LHC sensitivity extrapolated to 3/ab on the basis of the analyses provided in Tab.~\ref{tab:analyses}. Parameter points shown in green are allowed while point in red are excluded at 95\% confidence level by this analysis. In this particular figure, we suppress theoretical uncertainties,
but to reflect the impact of increased datasets on experimental systematics, we reduce the latter by 80\% which is provided 
by rescaling with the square root of the luminosity. A more detailed comparison of
experimental systematics and theoretical uncertainties is given in Fig.~\ref{fig:theoVSexpHLLHC}. \label{fig:scanHLLHC}}
\end{center}
\end{figure*}
%%%%%%%%%%%%%%%%%%%%%%%%%%%%%%%%%%

%%%%%%%%%%%%%%%%%%%%%%%%%%%%%%%%%%
\begin{figure*}[!t]
\begin{center}
\includegraphics[width=\textwidth]{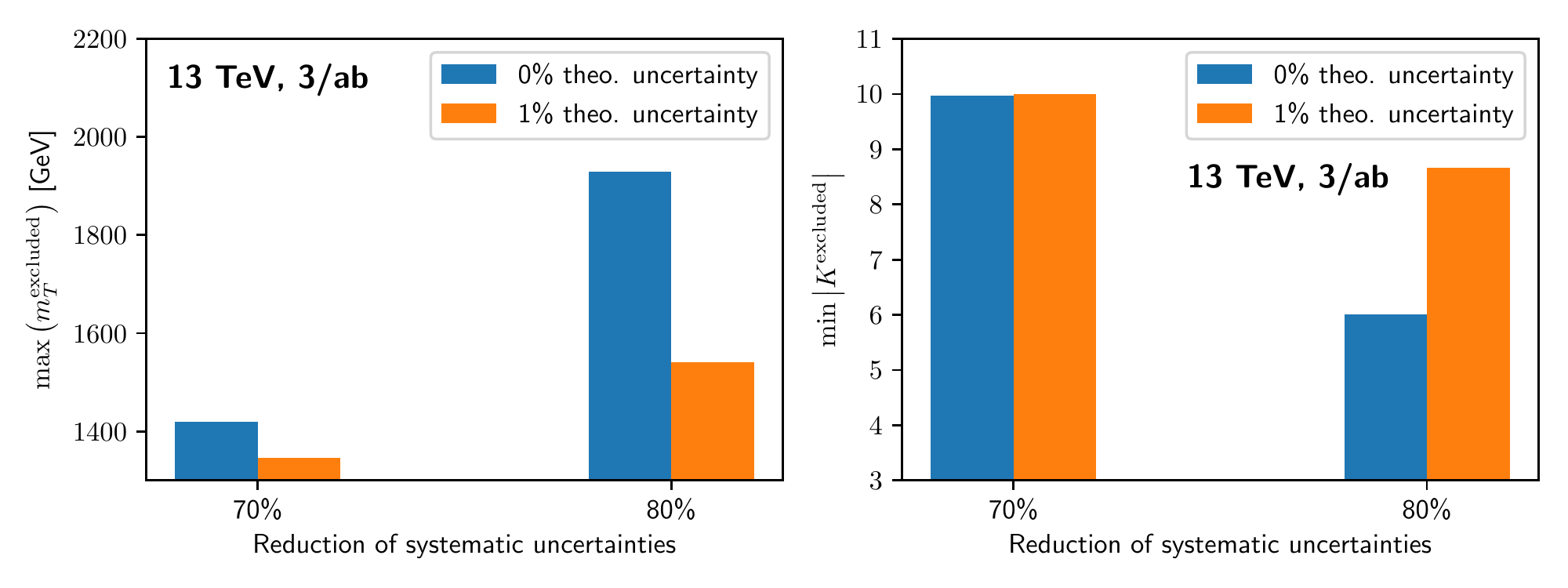}
\caption{Left: Maximum excluded top partner mass $m_T$ vs. reduction in experimental systematic uncertainties. The reduction is given with respect to the current experimental situation. The bars indicate different choices for relative theoretical uncertainties. Right: Minimal $|K|$ in the excluded region of the parameter scan vs. reduction in experimental systematic uncertainty.} 
\label{fig:theoVSexpHLLHC}
\end{center}
\end{figure*}
%%%%%%%%%%%%%%%%%%%%%%%%%%%%%%%%%%

%%%%%%%%%%%%%%%%%%%%%%%%%%%%%%%%%%
\section{Indirect Signs of Partial Compositeness: Present and High Energy Frontier}
\label{sec:results}
%%%%%%%%%%%%%%%%%%%%%%%%%%%%%%%%%%
Before we turn to the implications of the fit detailed in the previous section (and its extrapolations), we comment on additional constraints that could
be imposed from non-top data. 

Precision Higgs measurements are additional phenomenologically relevant channels that are sensitive to top partial compositeness through their modified Yukawa interactions. While the Yukawa sector probes different aspects of the model than the gauge interactions Eqs.~\eqref{eq:topmass} and \eqref{eq:jcurr}, they are equally impacted by the admixtures of vector-like top quarks, and are therefore correlated. For instance, the CMS projections provided in Ref.~\cite{CMS:2017cwx} can be used to comment on the relevance of the Higgs signal strength constraints: Out of all processes, $gg\to h, h\to ZZ$ provides the most stringent constraint when correlated with the top coupling deviations.\footnote{We note that derivative interactions $\sim K\, \bar t \gamma^\mu t \,\partial_\mu h$~\cite{Ferretti:2014qta} do not impact the loop-induced $h\to\gamma \gamma,gg$ amplitudes.}. The expected signal strength constraint at 3/ab of 4.7\% translates into a range of e.g. $|\dwl|\lesssim 0.18$. The 100 TeV extrapolation of Ref.~\cite{Benedikt:2018csr} of $\lesssim 2\%$ translates into $|\dwl|\lesssim 0.1$.

There are constraints from electroweak precision measurements, e.g.~\cite{Efrati:2015eaa}, which amount to a limit $|\delta g_{Z,L}| \lesssim 8\%$; flavour measurements provide an additional avenue to obtain limits on partial compositeness~\cite{Redi:2011zi,Redi:2012uj}. In the remainder, however, we focus on a comparison of direct top measurements at hadron colliders.

%%%%%%%%%%%%%%%%%%%%%%%%%%%%%%%%%%
\begin{figure*}[!p]
\begin{center}
\includegraphics[width=0.45\textwidth]{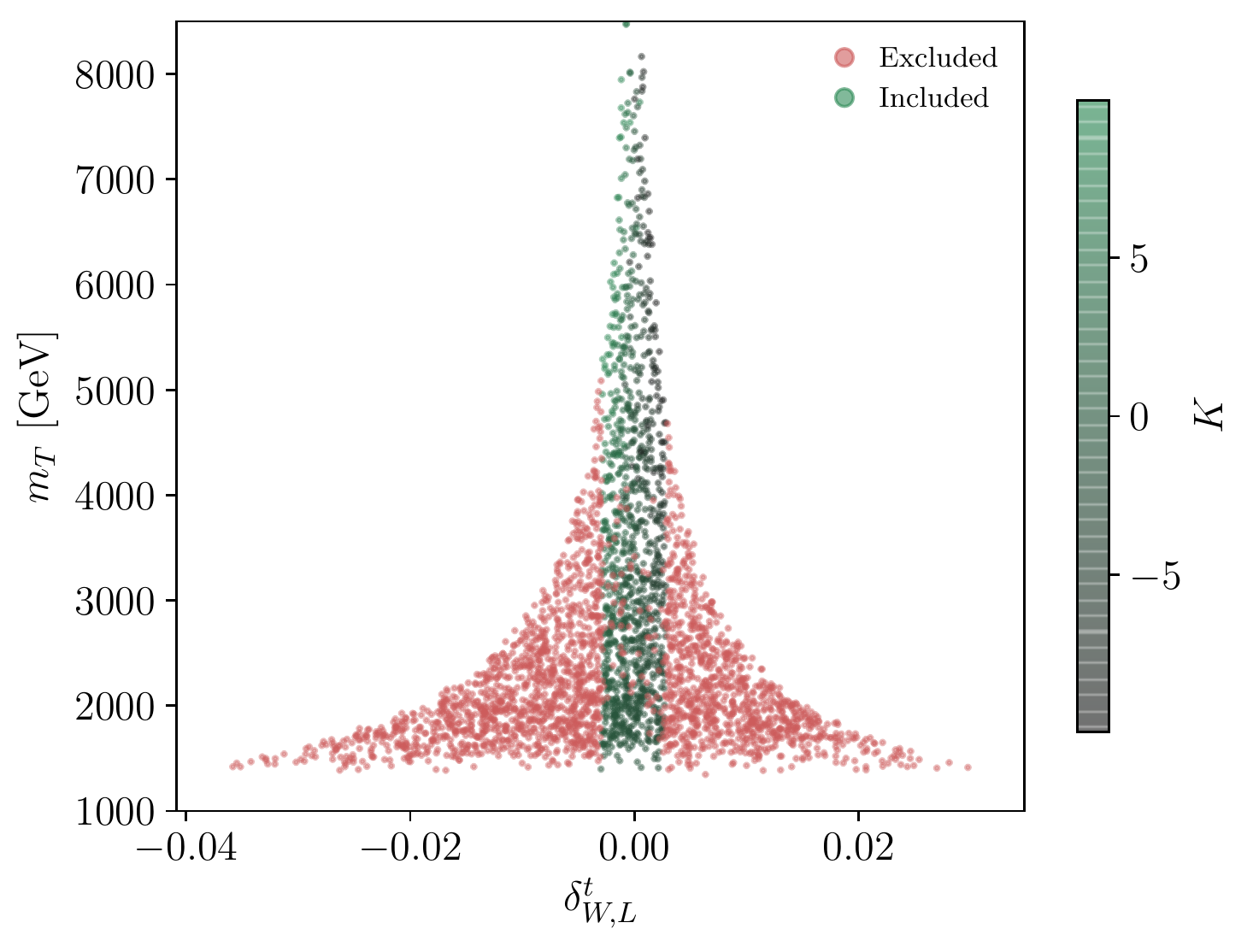}
\hskip 0.5cm
\includegraphics[width=0.45\textwidth]{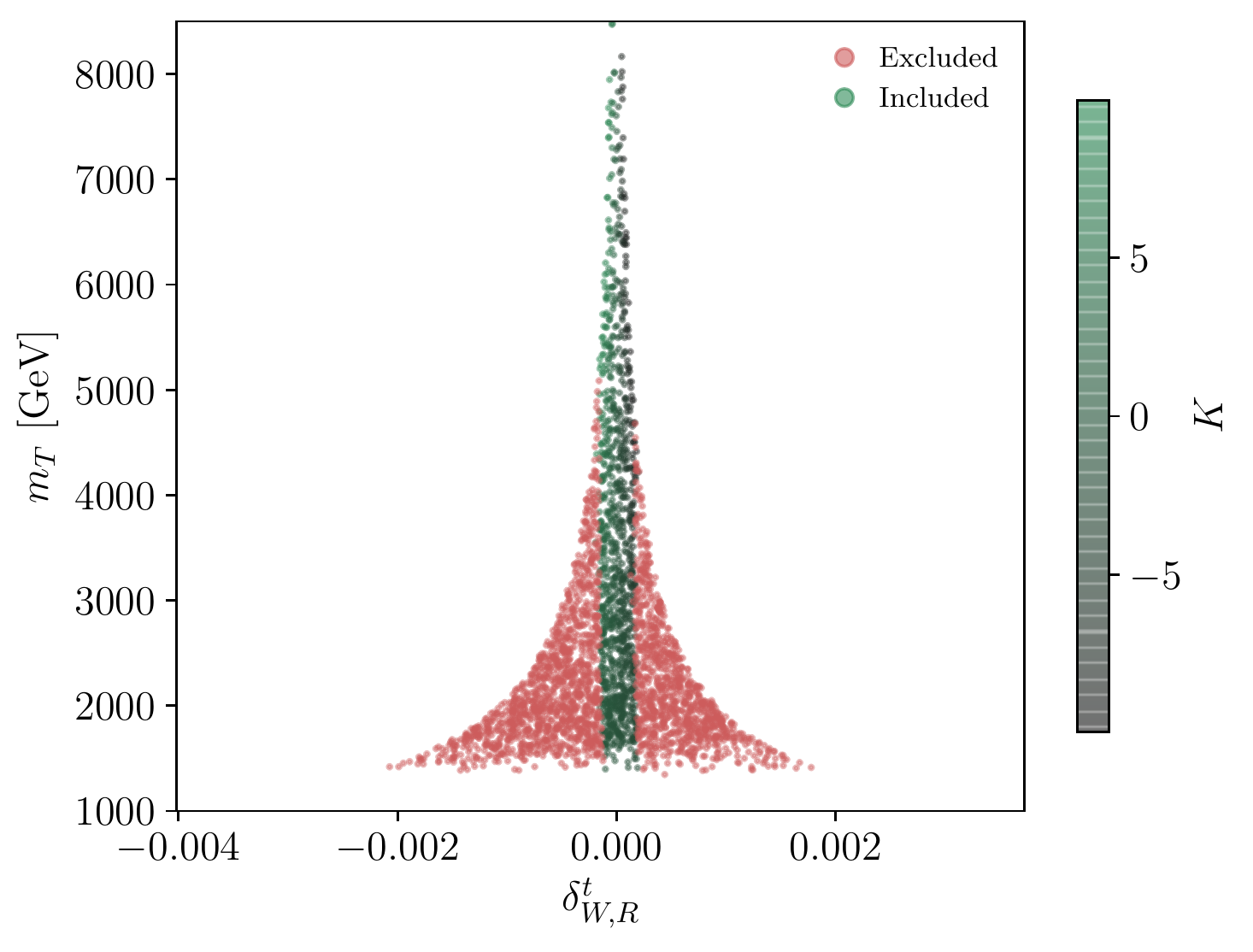}\\[0.3cm]
\includegraphics[width=0.45\textwidth]{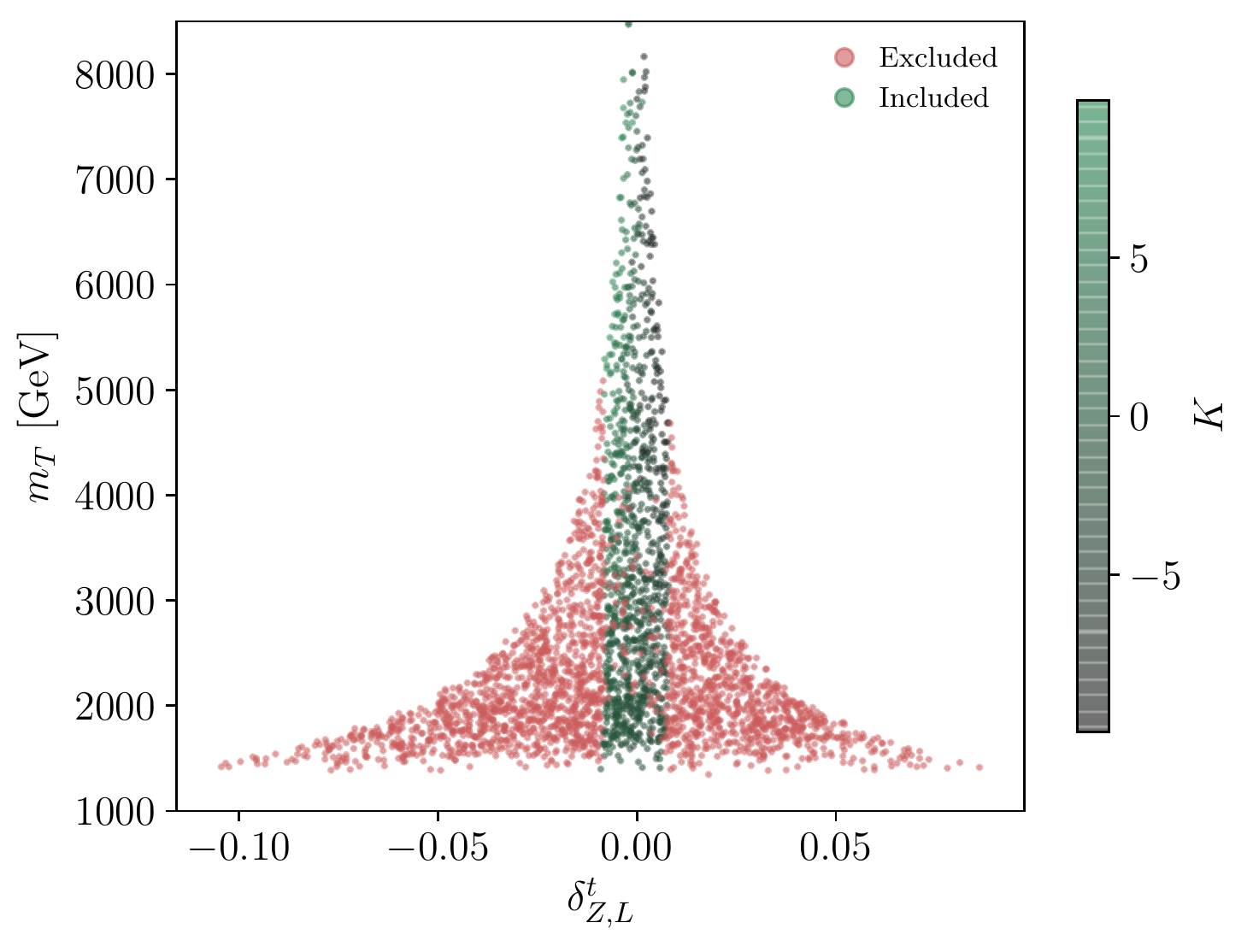}
\hskip 0.5cm
\includegraphics[width=0.45\textwidth]{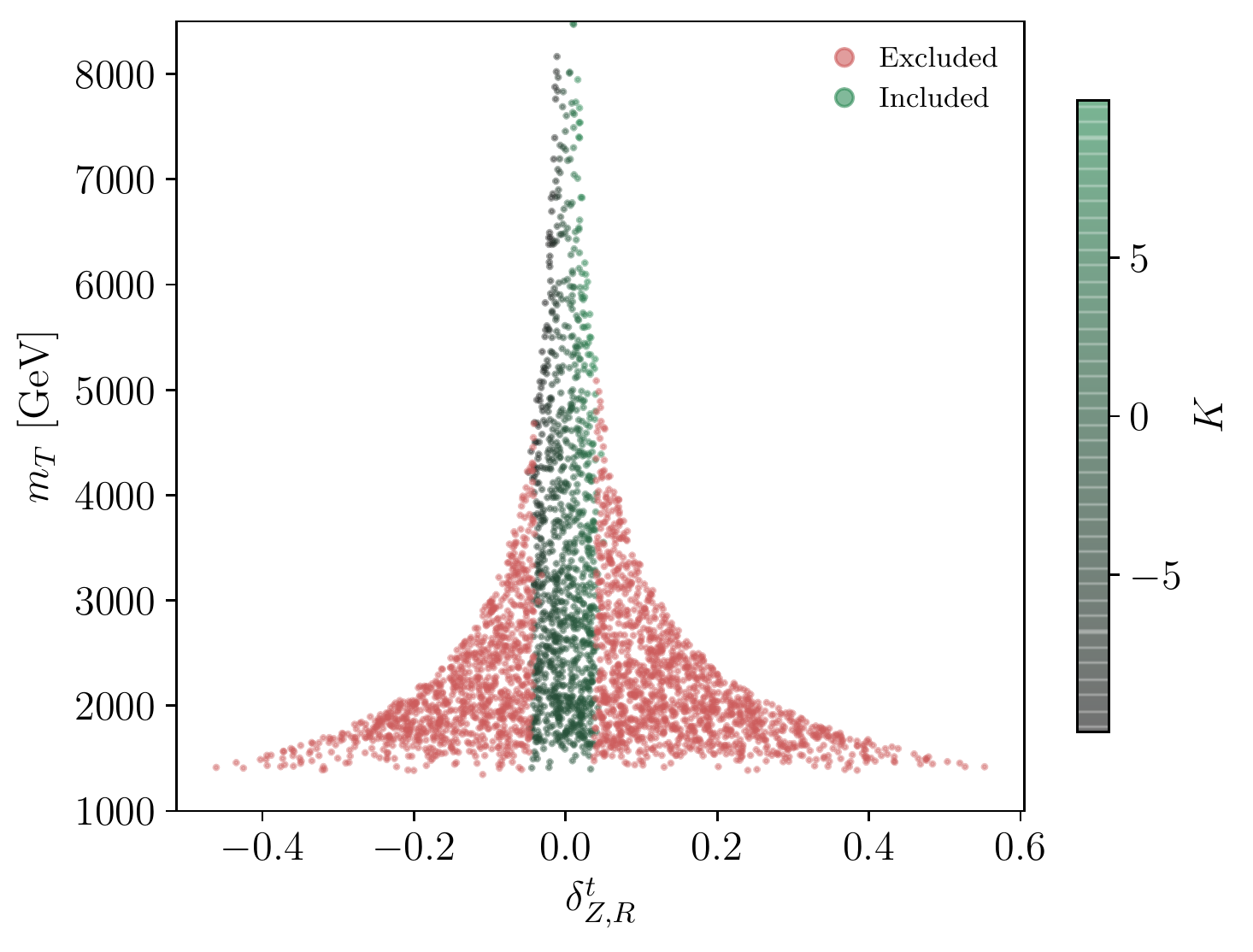}
\caption{Top coupling correlations analogous to Fig.~\ref{fig:scanHLLHC} for the FCC-hh analysis. We assume a reduction of experimental systematics to 1\% compared to the present LHC situation. In parallel, we suppress the theoretical uncertainty. See Fig.~\ref{fig:theoVSexpFCC} and the text for related discussion.\label{fig:scanFCC}}
\end{center}
\end{figure*}
%%%%%%%%%%%%%%%%%%%%%%%%%%%%%%%%%%
%%%%%%%%%%%%%%%%%%%%%%%%%%%%%%%%%%
\begin{figure*}[!p]
\begin{center}
\vfill
\includegraphics[width=\textwidth]{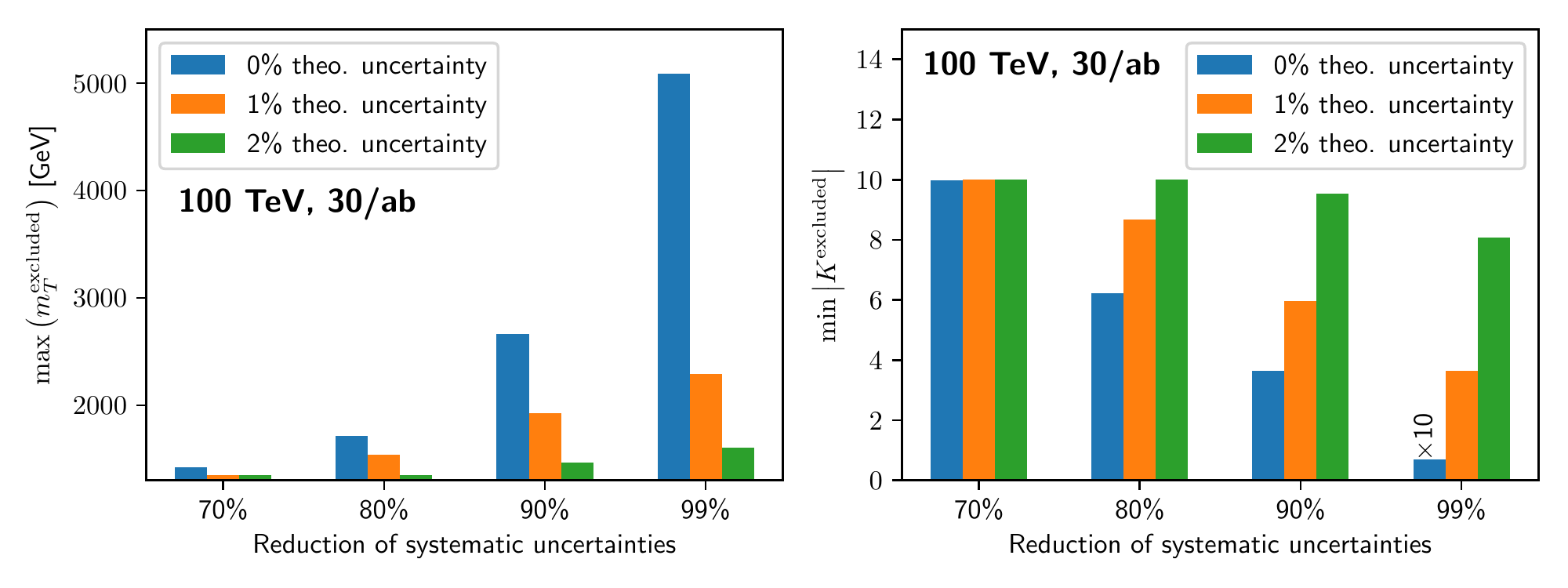}
\caption{Same as \Fig{fig:theoVSexpHLLHC} but for a centre-of-mass energy of $\sqrt{s}=100$ TeV and a luminosity of $L=30$/ab. The value of $\min|K^{\rm excluded}|$ for 99\% reduction in systematic uncertainties and no theory uncertainty was multiplied by a factor of 10 to increase visibility in the plot on the right-hand side.} 
\label{fig:theoVSexpFCC}
\end{center}
\end{figure*}
%%%%%%%%%%%%%%%%%%%%%%%%%%%%%%%%%%

As outlined in \Sec{sec:topew}, we scan over the parameters of the
Lagrangian in \Eq{eq:partcomp} imposing $M>1.5$~TeV to (loosely) reflect
existing top partner searches~\cite{Matsedonskyi:2012ym}. The restriction on
the parameter combination $\lambda_t\lambda_q$ is determined by $m_t\simeq
173$~GeV and on $\mu_b$ by the $b$ quark mass $m_b\simeq 4.7$~GeV (scanning
$|K|\lesssim 4\pi$). Apart from enforcing these masses we also consider
modifications to the Higgs boson decay and require the $H\to ZZ, \gamma\gamma$
decay rates to reproduce the SM predictions within 30\% to pre-select a reasonable parameter range. We fix the Higgs mass to 125 GeV as well as $v\simeq 246$~GeV in our scan, leaving $\xi$ (and hence $f$) as a free parameter. While the Higgs mass is directly linked with top and top partner spectra, we implicitly assume cancellations of the associated LEC parameters as expressed in Eq.~\eqref{eq:cancel} when taking into account top-partial compositeness.

We note that the degree of top compositeness is determined by the bi-unitary transformation of
Eq.~\eqref{eq:topmass}. In our scan, we find that the right-chiral top quark shows the largest degree of
compositeness, receiving 70\% to 90\% admixture from the hyperbaryon spectrum. In comparison, the left-chiral top is only $\lesssim 30\%$ composite in our scan. The right-chiral gauge coupling properties of the top are particularly relevant when we want to constrain this scenario, in particular given that they are absent in SM (see below).

Given the experimental results reported in \Tab{tab:analyses}, we find that the current
LHC (and Tevatron) measurements do not allow to constrain the parameter space detailed in
Sec.~\ref{sec:strongtop}
beyond the constraints that are already taken into account when
scanning the parameter space. Current Higgs signal constraints, for instance, 
provide stronger constraints. Since the top measurements are still at a relatively 
early stage in the LHC programme this is not too surprising, in particular because
top final states are phenomenologically more involved than their Higgs counterparts.

It is more interesting to consider how the sensitivity provided by the current
analysis programme of \Tab{tab:analyses} will evolve in the future. In \Fig{fig:scanHLLHC}, 
we present the results of the parameter scan for the
HL-LHC. The results are again based on the experimental
analyses in \Tab{tab:analyses} but with the statistical uncertainties rescaled
to 3/ab and experimental systematics reduced by 80\%.\footnote{This estimate is obtained from the statistical rescaling
\mbox{$\sim \sqrt{L_{\rm LHC}/L_{\rm HL-LHC}}\approx 0.2$} using the largest
so-far accumulated luminosity among the analyses in \Tab{tab:analyses}.} We assume
no theoretical uncertainties for now and will comment on their impact below. 
The observables of 7 and 8 TeV
analyses in \Tab{tab:analyses} are reproduced at 13 TeV\footnote{The total number of 
degrees of freedom for the projection of experimental data to $\sqrt{s}=13$ TeV and $L=3/$ab is $N=30$
due to the fact that we consider only one projection for each observable instead of several measurements.} keeping the
experimental bin-to-bin correlations of the respective analyses at their
original value.\footnote{We checked that the correlations have only a small effect on
the likelihood.} In \Fig{fig:scanHLLHC}, the excluded points of the parameter
scan are coloured in red while the allowed region is shaded in green. The shading
indicates the value of the parameter $K$.
As mentioned in \Sec{sec:strongtop}, the value of $K$ loosens the correlation between
the top partner mass and the associated electroweak top coupling modification. Furthermore, \Fig{fig:scanHLLHC} demonstrates 
that with higher luminosity and a (not unrealistic) reduction of the present systematic uncertainty we start to constrain
the parameter space with large $|K|\sim 10$ and associated coupling
deviations in the percent range, while the right-handed $Z$ coupling in the 30\% range.

In Fig.~\ref{fig:theoVSexpHLLHC} we compare different assumptions on the theoretical uncertainties in terms of
the maximal top partner mass $m_T$ and the minimal $|K|$ that can be excluded. Note that these are not strict
exclusion limits, smaller $m_T$ and larger $|K|$ might still be allowed. However, Fig.~\ref{fig:theoVSexpHLLHC} represents a measure of the
maximally possible sensitivity that can be probed at the HL-LHC in terms of the above quantities. As can be seen
in Fig.~\ref{fig:theoVSexpHLLHC}, the sensitivity of indirect searches crucially depends on the expected theoretical uncertainty
that will be achievable at the 3/ab stage. As for all channels that are not statistically limited at hadron colliders,
the theoretical error quickly becomes the limiting factor to the level where indirect searches will not provide
complementary information even at moderate top partner masses. 
A common practice \cite{Atlas:2019qfx, HLLHC:sysunc} for estimating projections for theoretical uncertainties at the HL-LHC is to apply a factor of 1/2 to the current theoretical uncertainties at the LHC. According to this prescription the projected theory uncertainties at the HL-LHC for for the observables studied in the analyses listed in \Tab{tab:analyses} are given by $\sim 1 - 5 \%$.

It is instructive to compare the approximate\footnote{Due to its granularity the scan provides only approximate bounds.} bounds on the anomalous couplings obtained in Fig.~\ref{fig:scanHLLHC}
\begin{equation*}
\begin{array}{lcllcl}
\dwl &\in& [-0.025,  0.02]  \,,\quad &
\dwr &\in& [-0.0014, 0.0013]\,,\\
\dzl &\in& [-0.073,  0.06]  \,,\quad &
\dzr &\in& [-0.33,   0.37]  
\end{array}
\end{equation*}
with 95\% CL marginalised limits obtained from a model
agnostic fit performed by \textsc{TopFitter} using the same experimental
projections
\begin{equation*}
\begin{array}{lcllcl}
\dwl &\in& [-0.029, 0.019]\,,\quad &
\dwr &\in& [-0.009, 0.009]\,,\\
\dzl &\in& [-0.639, 0.277]\,,\quad &
\dzr &\in& [-1.566, 1.350]\,.
\end{array}
\end{equation*}
In particular, the comparison of $\dwr$, $\dzl$, $\dzr$ between the two results
illustrates the fact that coupling deviations (or Wilson coefficients in the context of EFT)
are likely to receive much stronger constraints from the analyses of a concrete
model (possibly matched to EFT) due to correlations imposed by that model. This
highlights that recent multi-dimensional parameter fits~\cite{Buckley:2015lku,Castro:2016jjv,Durieux:2018tev,AguilarSaavedra:2018nen,Hartland:2019bjb,Brivio:2019ius,Durieux:2019rbz} are more sensitive to concrete realisations of high-scale new physics than the current model agnostic (marginalised) constraints might suggest. This will be further enhanced once we
move towards the high statistics realm of the LHC and whatever high energy frontier after that.

We now turn to the extrapolation of the analyses in Tab.~\ref{tab:analyses} to a future FCC-hh.
To this end we reproduce the observables in \Tab{tab:analyses} at a centre-of-mass energy of 100 TeV (we will comment on widening the list of observables below). In addition, we include overflow bins in $p_T$ distributions reflecting the fact that future analyses at 100 TeV will have a higher energy reach\footnote{The total number of degrees of freedom of the experimental results projected to $\sqrt{s}=100$ TeV and $L=30/$ab is $N=35$.}. In parallel, we rescale the statistical uncertainty from the analyses in \Tab{tab:analyses} to 30/ab and assume a reduction in systematic experimental uncertainties to 1\% of the LHC analyses.\footnote{Here we assume no theoretical uncertainty. A detailed comparison of the impact of uncertainties and experimental systematics is given in Fig.~\ref{fig:theoVSexpFCC}.}
For the 13 TeV analyses the bin-to-bin correlations have only a small impact on the exclusion of parameter points. Hence, we assume all measurements and bins in the 100 TeV analyses to be uncorrelated. The results for this scan are presented in \Fig{fig:scanFCC}, which shows that the FCC-hh can further improve on the LHC sensitivity 
by a factor of $\lesssim 3$ in terms of indirectly exploring the top partner mass in the scenario we consider in this work. 
Again theoretical uncertainties as parametrised in our scan are the key limiting factors of the sensitivity. There is no uniform convention or treatment for projecting theoretical uncertainties to the FCC-hh. However, at least with respect to QCD processes according to \Ref{Abada:2019lih} ``1\% is an ambitious but justified target''. In principle, a 100 TeV FCC-hh can reach $K={\cal{O}}(1)$ values as can be seen in Fig.~\ref{fig:theoVSexpFCC}. This is the perturbative parameter region where $T\rightarrow tZ$ direct searches (cf.~\cite{Reuter:2014iya}) are relevant. Hence, we focus on $|K|<1$ when we study this phenomenologically relevant channel in a representative top partner search in \Sec{sec:reson}.

Figs.~\ref{fig:theoVSexpHLLHC} and \ref{fig:theoVSexpFCC} demonstrate that the uncertainties as detailed in the previous section are the
key limiting factors of indirect BSM sensitivity in the near future. Naively, this paints a dire picture for the BSM potential. But we stress 
that data-driven approaches that have received considerable attention recently, e.g.~\cite{Aaboud:2018urx,Aad:2020kop}, together with the
application of new purpose-built statistical tools to mitigate the impact of uncertainties~\cite{Louppe:2016ylz,Brehmer:2018kdj,Brehmer:2018eca,Englert:2018cfo} will offer an avenue to inform constraints beyond ``traditional'' precision parton-level calculations at fixed order in perturbation theory. The basis of our analysis is also formed by extrapolating existing searches to 3/ab and eventually to 100 TeV. In particular, when statistics is not a limiting factor, a more fine-grained picture can be obtained by exploiting
differential information in more detail (see also a recent proposal to employ polarisation information in non-top channels~\cite{Cao:2020npb}). The latter, however, needs to be considered again in the context of experimental and theoretical limitations. Since the constraints on the $tZ$ coupling are the limiting factor in the indirect analysis considered here we have extended the inclusive $tjZ$ measurement by differential cross sections to assess the impact of additional differential information. To this end we include in the $tjZ$ channel the differential cross section with respect to the transverse momentum and the rapidity of the $Z$ boson. However, we do not find a significant change in the sensitivity projections as provided by Figs.~\ref{fig:theoVSexpHLLHC} and~\ref{fig:theoVSexpFCC}. A more detailed study of sensitive observables at hadron and lepton colliders is needed to maximise the sensitivity reach. But these excursions are beyond the scope of this work and are left for future studies.

%%%%%%%%%%%%%%%%%%%%%%%%%%%%%
\section{Top resonance searches}
\label{sec:reson}
%%%%%%%%%%%%%%%%%%%%%%%%%%%%%
The presence of additional vector-like fermions in composite Higgs models provides the opportunity of direct detection through resonance searches. We focus on channels involving the lightest top partner resonance (referred to as $T$ in the following) which can be either pair-produced through QCD interactions or created in association with a quark through interactions with vector bosons (or the Higgs boson). In particular, modes  $T \rightarrow t Z$, followed by decays of~$t Z \rightarrow ( q_1 q_2 b) (\ell^+ \ell^-)$ are interesting final states in the context of the previous section. On the one hand, they directly correlate modifications of electroweak top quark properties with new resonant structures following Eqs.~\eqref{eq:topmass} and \eqref{eq:jcurr}.
On the other hand, the presence of two same-flavour, oppositely charged leptons~$\ell^+$, $\ell^-$ (electrons or muons) in the boosted final state and no missing transverse energy allows discrimination between signal and background and the reconstruction of the top partner mass $m_T$ as demonstrated in Ref.~\cite{Reuter:2014iya}. We follow a similar cut-and-count analysis, adapted to FCC energies to attain a comparison with the indirect constraints of the previous section. Relevant SM background sources include $Z + $jets, $t \bar{t} Z + $jets and $t / \bar{t} Z + $jets, while the large mass of the top partner leading to a highly boosted $Z$ boson allows us to neglect the background processes involving two vector bosons and jets.

%%%%%%%%%%%%%%%%%%%%%%%%%%%%%%%%%%
\begin{figure*}[!t]
\begin{center}
\subfigure[~\label{fig:distrib}]{\includegraphics[width=0.45\textwidth]{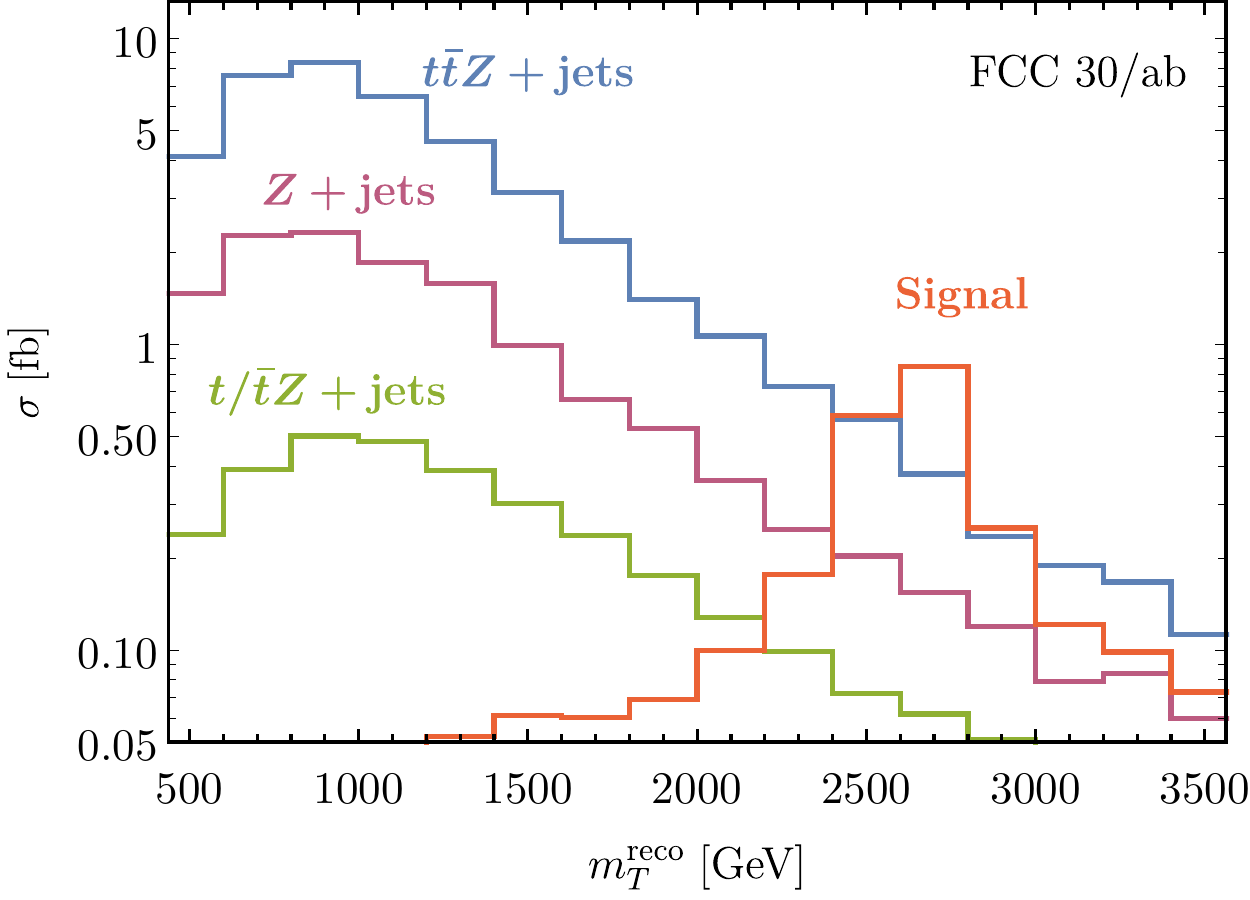}}
\hskip 0.5cm
\subfigure[~\label{fig:sensi}]{\includegraphics[width=0.45\textwidth]{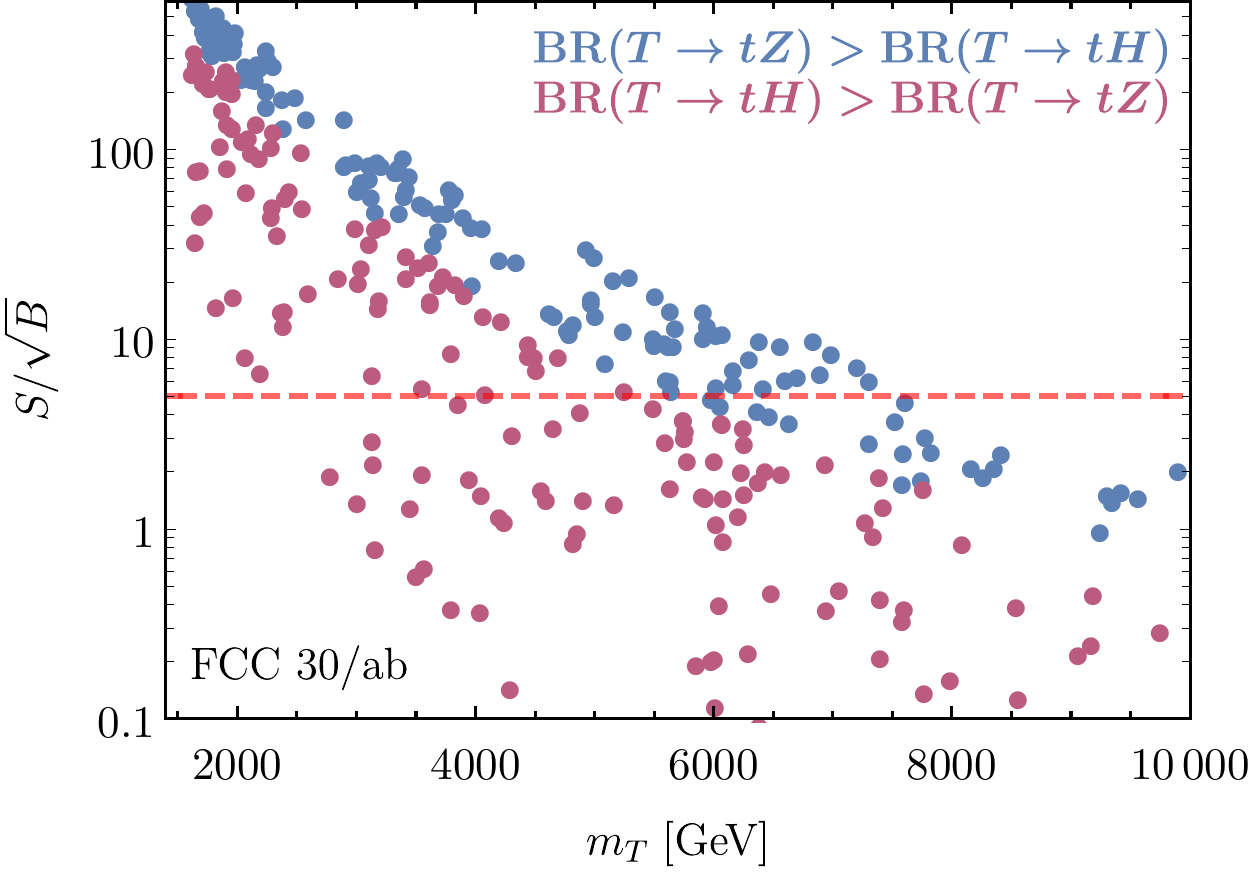}}
	\caption{(a) Differential cross sections for background and signal of a representative parameter point with a top partner mass of $m_T = 2700$~GeV. (b) Significance $S / \sqrt{B} $ for different coupling points at FCC $30$/ab is displayed on the right. The dashed red line indicates $S / \sqrt{B} = 5 $, where discovery can be achieved. For comparison, we include points dominantly decaying to $tH$ to show where our $tZ$ analysis is phenomenologically relevant.} 
	\label{fig:resonance_sign}
\end{center}
\end{figure*}
%%%%%%%%%%%%%%%%%%%%%%%%%%%%%%%%%%

We model the signal using {\sc{FeynRules}}~\cite{Christensen:2008py,Alloul:2013bka}, and events for both signal and background are generated with {\sc{MadEvent}}~\cite{Alwall:2011uj,deAquino:2011ub,Alwall:2014hca}. Decays are included via {\sc{MadSpin}}~\cite{Frixione:2007zp,Artoisenet:2012st} for the signal and $t \bar{t} Z + $jets,  $t / \bar{t} Z + $jets background processes. All events are showered with {\sc{Pythia8}}~\cite{Sjostrand:2014zea} using the {\sc{HepMC}} format~\cite{Dobbs:2001ck} before passing them to {\sc{Rivet}}~\cite{Buckley:2010ar} for a cut-and-count analysis, along with {\sc{FastJet}}~\cite{Cacciari:2011ma,Cacciari:2005hq} for jet clustering. The presence of a top in the boosted final state necessitates the use of jet-substructure methods for top-tagging, for which we adopt the Heidelberg-Eugene-Paris top-tagger ({\sc{HepTopTagger}})~\cite{Plehn:2010st,Kasieczka:2015jma,Plehn:2009rk}.

Final state leptons are required to be isolated\footnote{For a lepton to be isolated we require the total $p_T$ of charged particle candidates within the lepton's cone radius $\Delta R = 0.3$ to be less than $10 \%$ of the lepton's $p_T(\ell^\pm)$.} and have transverse momentum $p_T(\ell^\pm) \geq 20$~GeV and pseudorapidity $\abs{\eta(\ell^\pm)} \leq 2.5$. Slim-jets are clustered with the anti-kT algorithm~\cite{Cacciari:2008gp} with radius size of $0.4$ and fat-jets are also simultaneously reconstructed with Cambridge-Aachen algorithm and a larger size of $1.5$. Both types of jets must satisfy $p_T(j) \geq 20$~GeV and $\abs{\eta(j)} \leq 4.9$. 

Lepton selection cuts are applied by requiring at least one pair of same flavour oppositely charged leptons, with an invariant mass within $10$~GeV of the $Z$ boson resonance, i.e. $\abs{ m_{\ell^+ \ell^-} - m_Z} < 10$~GeV. Furthermore, we require $\Delta R(\ell^+ \ell^-) = \sqrt{[\Delta \eta(\ell^+ \ell^-) ]^2 + [\Delta \phi(\ell^+ \ell^-) ]^2} < 1.0$ to ensure that the leptons are collimated. The two leptons must have a minimum transverse momentum of $p_T(\ell^\pm) > 25$~GeV, and if more than one candidate pairs exist, the one with invariant mass closest to $m_Z=91.1~\text{GeV}$ is selected to reconstruct the $Z$ boson's four-momentum. Subsequently, the search region is further constrained with the requirements $p_T(Z) > 225$~GeV and $\abs{\eta(Z)} < 2.3$, where the former further ensures the boosted kinematics and the latter allows better discrimination from the $Z + $jets background of the SM.

The hadronic part of the signal's final state is characterised by large transverse momentum originating from the top quark's boosted nature and thus we require that the scalar sum of the transverse momenta satisfies $H_T > 700$~GeV for all identified slim-jets that have $p_T(j) > 30$~GeV and $\abs{\eta(j)} < 3$. The search region is constrained by requiring at least one fat-jet that satisfies $p_T(j) > 200$~GeV and is top-tagged with {\sc{HepTopTagger}}. In the case of more than one top candidate we consider the one where $\Delta \phi ( Z , t )$ is closest to $\pi$, ensuring the $Z$ and $t$ candidates are back-to-back. B-jets are identified from slim-jets and at least one satisfying $p_T(b) > 40$~GeV is required to be within the top radius of $\Delta R (t , b) < 0.8$, implying that the b quark originated from the top. The b-tag efficiency is set to $80\%$, while the mistagging probability of quarks at $1\%$. Finally, the reconstructed top and $Z$ candidates are used to reconstruct the top partner's mass $m_T^{\text{reco}}$ via the sum of the $Z$ and $t$ four momenta.

The efficiency of the cut-and-count analysis is determined by the resonance mass, which defines the kinematics of the final state 
particles. We scan over a range of top partner masses and perform an interpolation to eventually evaluate constraints in a fast and adapted way. We have validated the accuracy of this approach against additional points as well as against the independence of the coupling values. We find that a signal region definition using the reconstructed top partner mass $m_T^{\text{reco}} \in \left[ m_T - 0.2 m_T , m_T + 0.15 m_T \right]$ is an appropriate choice to reduce backgrounds and retain enough signal events to
set limits in the region $|K|<1$ that we are interested in as detailed before. This ensures that the detailed search is perturbatively under control and phenomenologically relevant. For larger $K$ values the $T\rightarrow ht$ decay receives sizeable momentum-dependent corrections~\cite{Ferretti:2014qta}, which quickly start to dominate the total decay width to a level where we can expect our analysis flow to become challenged due to non-perturbative parameter choices.

In the spirit of data-driven bump hunt searches we fit the $m^{\text{reco}}_T$ distribution away from the signal region to 
obtain a background estimate in the signal region defined above. As can be seen in Fig.~\ref{fig:distrib}, such distributions
follow polynomial distributions on a logarithmic scale and are therefore rather straightforward to control in a data-driven approach. 
There we show a $m_T^\text{reco}$ histogram for a representative signal point $m_T\simeq 2.7$~TeV and the contributing background. 
Such a data-driven strategy also largely removes the influence of theoretical uncertainties at large momentum transfers and is the typical method of
choice in actual experimental analyses already now, see e.g. \cite{Aaboud:2018urx,Aad:2020kop} for recent work. After all analysis steps are carried out we typically deal with a signal-to-background ratio $S/B\sim 0.1$, which means that our sensitivity is also not too limited by the background uncertainty that would result from such a fit.
Identifying a resonance, we can evaluate the significance which is controlled by $S / \sqrt{B}$. To set limits we assume a total integrated luminosity of 30/ab for 100 TeV FCC-hh collisions. We show sensitivity projections in \Fig{fig:sensi}. As can be seen we have good discovery potential in $tZ$ for parameter regions up to $m_T\simeq 7.3$~TeV, with the additional exclusion potential $\sim S/\sqrt{S+B}$ reaching to $m_T \lesssim 10$~TeV at 95\% CL. As alluded to before, the analysis outlined above is particularly suited for parameter regions where there is a significant top partner decay into $Zt$ pair, i.e. regions in parameter space where modifications are most pronounced in the weak boson phenomenology rather than in Higgs-associated channels. 

While we have focused on one particular analysis to contextualise the couplings scan of the previous section with representative direct sensitivity at the highest energies, we note that other channels will be able to add significant BSM discovery potential, see, e.g. Refs.~\cite{Golling:2016gvc,Matsedonskyi:2014mna}. 
This could include $T\to ht$ which would lead to $b$-rich final states and which would target partial compositeness in the Higgs sector (see also~\cite{Barducci:2017ddn,Li:2019ghf}). Such an analysis provides an avenue to clarify the Higgs sector's role analogous to the weak boson phenomenology studied in this work, albeit in phenomenologically more complicated final states when turning away from indirect Higgs precision analyses and $t\bar t h$ production. Furthermore searches for other exotic fermion resonances different to the one we have focused on in this section, such $B$ and the $5/3$-charged $Q$ provide additional discriminating power (see~\cite{Azatov:2013hya,Sirunyan:2018yun}) and would be key to pinning down the parameter region of the model if a new physics discovery consistent with partial compositeness is made. 

\bigskip
Being able to finally compare the direct sensitivity estimates of Fig.~\ref{fig:resonance_sign} with Fig.~\ref{fig:scanFCC} we see that indirect searches for top compositeness as expressed through modifications of the top's SM electroweak couplings provide additional information to resonance searches if uncertainties can be brought under sufficient control. For instance, the potential discovery of the top partner alone is insufficient to verify or falsify the model studied in this work. The correlated information of top quark coupling deviations is an additional crucial step in clarifying the underlying UV theory.

Extrapolating the current sensitivity estimates of the LHC alongside the uncertainties to the 3/ab phase, the HL-LHC will however provide only limited insight from a measurement of the top's electroweak SM gauge interaction deformations. This can nonetheless lead to an interesting opportunity at the LHC: Given that the LHC will obtain a significantly larger sensitivity via direct searches \cite{Azatov:2013hya,Reuter:2014iya,Sirunyan:2018yun}, the potential discovery of a top partner at the LHC would make a clear case for pushing the energy frontier to explore the full composite spectrum and correlate these findings with an enhanced sensitivity to top coupling modifications.

%%%%%%%%%%%%%%%%%%%%%%%%%%%%%
\section{Conclusions}
\label{sec:conc}
%%%%%%%%%%%%%%%%%%%%%%%%%%%%%
As top quark processes can be explored at the LHC with high statistics, they act as Standard Model ``candles''. The electroweak properties of the top quark are particularly relevant interactions as deviations from the SM are tell-tale signatures of new physics beyond the SM that is directly relevant for the nature of the TeV scale.

Using the example of top partial compositeness (and the extended MCHM5 implementation of \cite{Ferretti:2014qta} for concreteness) we demonstrate that the ongoing top EFT programme will provide important additional information to resonance searches if theoretical and experimental uncertainties will be brought under control. This is further highlighted at the energy frontier of a future hadron collider at 100 TeV. Backing up our electroweak top coupling analysis with a representative top partner resonance search, we demonstrate the increased sensitivity and additional discriminating power to pin down the top quark's electroweak properties at the FCC-hh. Especially in case a discovery is made at the LHC that might act as a harbinger of a composite TeV scale, there is a clear case for further honing the sensitivity to the top's coupling properties whilst extending the available energy coverage. We note that high-energy lepton colliders such as CLIC will
be able to provide a very fine grained picture of the top electroweak interactions, which can provide competitive indirect sensitivity~\cite{Englert:2017dev,Durieux:2018tev,Escamilla:2017pvd,vanderKolk:2017diw,Boronat:2019cgt,Abramowicz:2018rjq}. We leave a more detailed comparison of the interplay of hadron and lepton colliders for future work.

\acknowledgments
%\bigskip\noindent{\bf{Acknowledgements}} ---
We thank Federica Fabbri for helpful discussions.
SB is funded by the UK Science and Technology Facilities Council (STFC) through a ScotDIST studentship under grant ST/P006809/1.
CE and PG are supported by the STFC under grant
ST/P000746/1. CE also acknowledges support through the IPPP associate scheme.
PS is supported by an STFC studentship under grant ST/T506102/1.

%%%%%%%%%%%%%%%%%%%%%%%%%%%%%%%%%%
\appendix
%%%%%%%%%%%%%%%%%%%%%%%%%%%%%%%%%%
%%%%%%%%%%%%%%%%%%%%%%%%%%%%%%%%%%
\section{Propagating top partners as EFT contributions}
\label{app:eft}
%%%%%%%%%%%%%%%%%%%%%%%%%%%%%%%%%%
On top of the coupling modifications of the top-associated currents, amplitudes receive corrections
from propagating top partners. Similarly a composite top substructure can lead to additional anomalous
magnetic moments \cite{Englert:2012by,Barducci:2017ddn,BuarqueFranzosi:2019dwg} as observed in
nuclear physics~\cite{Holstein:2000ap}. At the considered order in the chiral expansion in this work such 
terms arise at loop level~\cite{Schwinger:1948iu,Jersak:1981sp}, and at tree level 
via the direct propagation of top partners. It is interesting to understand the latter contributions from 
an EFT perspective as they not only give rise to dimension six effects and cancellations can occur. 
In the mass eigenbasis, the propagating degrees of freedom lead to dimension
eight effects. For instance, $t\bar t \to WW$ scattering in the mass eigenbasis receives corrections from $b,B$
as well as from the 5/3-charged $Q$. The resulting Lorentz structure of contact $t\bar t W^+W^-$ amplitude in the EFT-limit is described by a combination
of 
\begin{equation}
\begin{split}
{\cal{O}}_{tW} &=  \bar{Q}_L \sigma^{\mu\nu} \tilde \varphi \, \tau^a t_R W^{a}_{\mu\nu}\\
{\cal{O}}_{tH} &=  (D_\mu \varphi^\dagger D^\mu \varphi) \bar{Q}_L \tilde \varphi\, t_R  
\end{split}
\end{equation}
leading to
\begin{equation}
{\cal{M}}(t\bar t \to W^+W^-) = {C_{tW}\over \Lambda^2} \langle {\cal{O}}_{tW} \rangle +    {C_{tH}\over \Lambda^4} \langle {\cal{O}}_{th} \rangle + \dots
\end{equation}
where the ellipses refer to momentum-dependent corrections that become relevant for $Q^2\sim m_X^2$.
\begin{equation}
\begin{split}
  {C_{tW}\over \Lambda^2} &= -{g_W\over 4 m_t}  \left( {c^{tB}_Lc^{tB}_R\over m_B} -{ c^{tX}_L c^{tX}_R\over m_X}  \right)\,,\\
  {C_{tH}\over \Lambda^4} &= -{g_W^2\over 16 m_tm_W^2 }  \left( {c^{tB}_Lc^{tB}_R\over m_B} +{ c^{tX}_Lc^{tX}_R\over m_X}  \right)\,,
\end{split}
\end{equation}
where the $e\,c^{tX}_{L,R},e\,c^{tB}_{L,R}$ are the left and right-chiral $W$ couplings of the top with the respective top partner in the mass basis.

%%%%%%%%%%%%%%%%%%%%%%%%%%%%%%%%%%
\section{EFT parametrisation of anomalous weak top quark couplings}
\label{app:anomalous}
%%%%%%%%%%%%%%%%%%%%%%%%%%%%%%%%%%
The effective dimension six operators (in the Warsaw basis~\cite{Grzadkowski:2010es}) that modify the
vectorial couplings of the top quark to the $W$ and $Z$ bosons are given by
\begin{equation}
\begin{split}
\OHq{(1)} & =  (\vpj)(\bar Q \gamma^\mu Q)\,,\\
\OHq{(3)} & =  (\vpjt)(\bar Q \tau^I \gamma^\mu Q)\,,\\
\OHu      & =  (\vpj)(\bar t_R \gamma^\mu t_R)\,,\\
\OHud     & =  i(\tvp^\dag D_\mu \vp)(\bar t_R \gamma^\mu b_R)\,,\\[0.4cm]
\end{split}
\end{equation}
with the associated Wilson coefficients $\CHq{(1)}$, $\CHq{(3)}$, $\CHu$ and $\CHud$.
See also Ref. \cite{Cao:2015doa} for a detailed recent discussion beyond tree-level.
$Q=(t_L, b_L)^T$ denotes the quark $SU(2)_L$ doublet of the third generation with 
$t_L$ and $b_L$ the left-handed top and bottom quarks, respectively. The rest of the notation
is aligned with \Ref{Grzadkowski:2010es}. 
The anomalous couplings of the top quark to $W$ and $Z$ bosons are related to the
Wilson coefficients as follows
\begin{subequations}
\label{eq:deltacoup}
\begin{align}
\dzl & =  -\frac{\CHqZ\vev^2}{\Lambda^2}\left(1-\frac{4}{3}\sin^2\theta_W\right)^{-1}\,,
\label{eq:deltaZtL}\\
\dzr & =  \frac{\CHu\vev^2}{\Lambda^2}\frac{3}{4\sin^2\theta_W}\,,
\label{eq:deltaZtR}\\
\dwl   & =  \frac{\CHqW\vev^2}{\Lambda^2}\,,
\label{eq:deltaWL}\\
\dwr   & =  -\frac{1}{2}\frac{\CHud\vev^2}{\Lambda^2}\,.
\label{eq:deltaWR}
\end{align}
\end{subequations}
In Eqs.~\eqref{eq:deltaZtL} and \eqref{eq:deltaWL} we have introduced two new Wilson
coefficient which correspond to the operators
\begin{equation}
\begin{split}
\OHqW & =  \OHq{(3)}\,,\\
\OHqZ & =  \OHq{(1)} - \OHq{(3)}\,.
\end{split}
\end{equation}
This change of basis ensures that each of the four operators $\OHqW$, $\OHqZ$,
$\OHu$ and $\OHud$ contributes to exactly one kind of $W$ and $Z$ coupling in
\Eq{eq:WZcoupling}.  The relations of Eq.~\eqref{eq:deltacoup} allow us to
directly relate constraints on the Wilson coefficients to constraints on the
coupling modifications~$\delta$.

%%%%%%%%%%%%%%%%%%%%%%%%%%%%%%%%%%
%\bibliographystyle{apsrev4-1}
\bibliography{references} 
%\bibliography{paper.bbl}

\end{document}